\documentclass[10pt,twocolumn,letterpaper]{article}

\usepackage[pagenumbers]{cvpr} 
\usepackage{threeparttable}
\usepackage{float}
\usepackage{pifont}
\usepackage{footnote}
\usepackage{enumitem}
\usepackage{bm}
\usepackage{arydshln}
\usepackage{booktabs}
\usepackage{multicol}
\usepackage{multirow}
\usepackage{color}
\usepackage{xcolor}     
\usepackage{colortbl}
\usepackage{soul}
\usepackage{bbding}
\usepackage{makecell}
\usepackage{mathtools}
\usepackage{imakeidx}
\makeindex
\usepackage{arydshln}
\usepackage{lipsum}
\usepackage{natbib}
\usepackage[toc]{multitoc}
\usepackage[edges]{forest}
\usepackage[normalem]{ulem}

\usepackage{bbding}
\usepackage[most]{tcolorbox}

\usepackage{algorithm}
\usepackage{algorithmic}

\usepackage{minitoc}
\usepackage[toc,page,header]{appendix}

\definecolor{orchid}{rgb}{0.85, 0.44, 0.84}

\newcommand{\method}{{\fontfamily{ppl}\selectfont
TextoMorph}}

\newcolumntype{P}[1]{>{\centering\arraybackslash}p{#1}}
\newlength\savewidth

\definecolor{cvprblue}{rgb}{0.21,0.49,0.74}
\usepackage[pagebackref,breaklinks,colorlinks,allcolors=cvprblue]{hyperref}

\def\confName{CVPR}
\def\confYear{2025}

\title{Text-Driven Tumor Synthesis}

\author{Xinran Li\textsuperscript{1,2} \quad 
Yi Shuai\textsuperscript{3,4} \quad 
Chen Liu\textsuperscript{1,5} \quad 
Qi Chen\textsuperscript{1,6} \quad
Qilong Wu\textsuperscript{1,7} \quad
Pengfei Guo\textsuperscript{8} \quad 
Dong Yang\textsuperscript{8} \quad \\
Can Zhao\textsuperscript{8} \quad 
Pedro R. A. S. Bassi\textsuperscript{1,9,10} \quad   
Daguang Xu\textsuperscript{8} \quad 
Kang Wang\textsuperscript{11} \quad 
Yang Yang\textsuperscript{11} \\ 
Alan Yuille\textsuperscript{1} \quad 
Zongwei Zhou\textsuperscript{1,}\thanks{Correspondence to Zongwei Zhou (\href{mailto:zzhou82@jh.edu}{\textsc{zzhou82@jh.edu}})} \\[2.5mm]
\textsuperscript{1}Johns Hopkins University \quad
\textsuperscript{2}Shenzhen Technology University \quad
\textsuperscript{3}Sun Yat-sen University \\
\textsuperscript{4}The First Affiliated Hospital of Sun Yat-sen University\quad
\textsuperscript{5}Hong Kong Polytechnic University \\
\textsuperscript{6}University of Chinese Academy of Sciences \quad
\textsuperscript{7}National University of Singapore \quad
\textsuperscript{8}NVIDIA \\
\textsuperscript{9}University of Bologna \quad
\textsuperscript{10}Italian Institute of Technology \quad
\textsuperscript{11}University of California, San Francisco \\[1.5mm]
{\small Code, dataset, and models:~\href{https://github.com/MrGiovanni/TextoMorph}{https://github.com/MrGiovanni/TextoMorph}}
}

\begin{document}
\maketitle

\doparttoc 
\faketableofcontents 
\begin{abstract}

    Tumor synthesis can generate examples that AI often misses or over-detects, improving AI performance by training on these challenging cases. However, existing synthesis methods, which are typically unconditional---generating images from random variables---or conditioned only by tumor shapes, lack controllability over specific tumor characteristics such as texture, heterogeneity, boundaries, and pathology type. As a result, the generated tumors may be overly similar or duplicates of existing training data, failing to effectively address AI's weaknesses.

    We propose a new text-driven tumor synthesis approach, termed \textbf{\method}, that provides textual control over tumor characteristics. This is particularly beneficial for examples that confuse the AI the most, such as early tumor detection (increasing Sensitivity by \textbf{+8.5\%}), tumor segmentation for precise radiotherapy (increasing DSC by \textbf{+6.3\%}), and classification between benign and malignant tumors (improving Sensitivity by \textbf{+8.2\%}). By incorporating text mined from radiology reports into the synthesis process, we increase the variability and controllability of the synthetic tumors to target AI's failure cases more precisely. Moreover, \method\ uses contrastive learning across different texts and CT scans, significantly reducing dependence on scarce image-report pairs (only 141 pairs used in this study) by leveraging a large corpus of 34,035 radiology reports. Finally, we have developed rigorous tests to evaluate synthetic tumors, including Text-Driven Visual Turing Test and Radiomics Pattern Analysis, showing that our synthetic tumors is realistic and diverse in texture, heterogeneity, boundaries, and pathology.

\end{abstract}

\section{Introduction}\label{sec:introduction}

\begin{figure}[t]
    \centering
    \includegraphics[width=\linewidth]{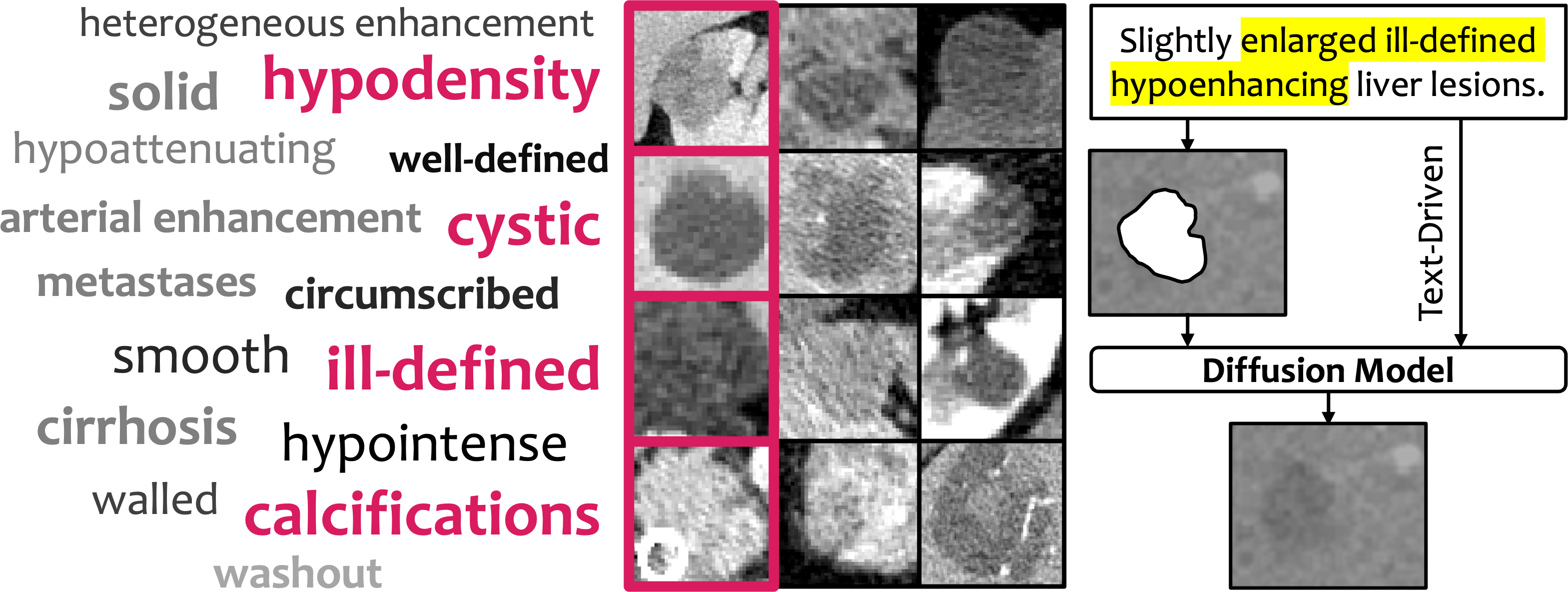}
    \caption{\textbf{Text-Driven Tumor Synthesis.} Existing tumor synthesis methods struggle with limited controllability, often generating tumors based solely on predefined shapes or random noise. This results in synthetic data that lacks essential features like texture, boundaries, and attenuation, reducing its effectiveness in addressing AI weaknesses. \textbf{\method} addresses this limitation by exploiting a dataset of 34,176 radiology reports to generate tumors with medically precise features described in clinical language. Examples include phrases such as `\textit{hypodensity}', `\textit{ill-defined}', and `\textit{cystic}', paired with CT scans of the liver, pancreas, and kidney.}
    \label{fig:ct_report}
\end{figure}

Tumor synthesis plays a critical role in \textit{targeted} data augmentation by generating examples that AI models tend to miss (false negatives) or over-detect (false positives), focusing on areas needing improvement \cite{niemeijer2024tsynd,basaran2023lesionmix} and addressing privacy concerns and reducing annotation costs \cite{chen2024analyzing,hu2023label,lai2024pixel}. However, existing synthesis methods are typically unconditional \cite{gonccalves2024abdominal}---generating images from random variables---or conditioned only on shape masks \cite{chen2024towards}, lacking controls over specific tumor characteristics such as texture, heterogeneity, boundaries, and pathology type. We find that text should be considered an important conditioning factor when generating tumors because it carries much richer information\footnote{For example, a report goes `\textit{slightly enlarged ill-defined liver lesions}' and `\textit{more well-defined appearance of liver lesions}.' The corresponding CT scans are shown in \figureautorefname~\ref{fig:ct_report}.} than random variables or shape masks can offer. Moreover, tumor-related text is readily available in radiology reports, which are routinely generated by radiologists in clinical workflows. We hypothesize that incorporating text as a condition, alongside tumor masks, allows us to develop stronger AI models for tumor detection, segmentation, and classification due to greater controllability of generating such tumors that AI models often make mistakes.

Text-driven generative models \cite{xu2018attngan,saharia2022photorealistic,nichol2021glide} have significantly advanced in recent years. These models leverage natural language descriptions to control the synthesis of images/videos, enabling fine-grained manipulation of generated content. Applications range from data augmentation for AI training to commercial products that generate images/videos for creative and practical purposes. However, these models have not been fully explored in tumor synthesis due to several challenges: 
\textit{\textbf{First}, lack of annotated tumor images:} Only a very small proportion (less than 5\%) of publicly available abdominal CT datasets contain annotated tumors \cite{bilic2019liver,heller2023kits21,roth2015deeporgan,chou2024embracing,bassi2024touchstone}.
\textit{\textbf{Second}, lack of text descriptions:} None of the publicly available abdominal CT scans have paired radiology reports or text descriptions.
\textit{\textbf{Third}, need of large-scale paired datasets for training:} For example, DALL·E was trained on 250 million image-text pairs \cite{ramesh2021zero}, and Imagen Video was trained on 14 million video-text pairs along with 60 million image-text pairs \cite{ho2022imagen}.
\textit{\textbf{Forth}, difficulty in evaluating generated synthetic tumors:} AI-generated images/videos can be assessed by anybody, while generated tumors must be visually inspected by busy, costly medical professionals \cite{hu2022synthetic,du2024boosting,li2023early,hu2023label,xu2024medsyn}.

To address these challenges, we first create a dataset consisting of 141 CT-Report pairs containing tumors in the liver, pancreas, and kidney, along with 34,035 radiology reports that provide textual descriptions of tumors or normal findings (see examples in \figureautorefname~\ref{fig:ct_report}). We then develop a new text-driven tumor synthesis approach, termed \textbf{\method}, which can generate targeted tumors based on the described tumor characteristics. By incorporating textual descriptions mined from radiology reports into the synthesis process, \method\ increases the variability and controllability of the synthetic tumors, allowing us to precisely target the AI's failure modes. This is particularly beneficial for such cases that challenge AI the most, including (1) early-stage tumor detection (less than 20mm), increasing Sensitivity by \textbf{+8.5\%} (\appendixautorefname~\ref{sec:supp_Ablation_overall}), (2) tumor segmentation for precise radiotherapy, increasing DSC by \textbf{+6.3\%} (\appendixautorefname~\ref{sec:supp_Ablation_overall}), and (3) classification between benign and malignant tumors, improving Sensitivity by \textbf{+8.2\%} (\tableautorefname~\ref{tab:tumor_classification}). More importantly, we have also developed rigorous tests to evaluate the effectiveness of synthetic tumors for targeted data augmentation. 
\textit{\textbf{First}, \ul{Text-Driven Visual Turing Test} to examine tumor fidelity.} Radiologists were asked to distinguish real and synthetic tumors with the same shape mask and text description (e.g., both being cystic tumors). As shown in \tableautorefname~\ref{tab:visual_turing_test}, they erred 22.5--45.0\% of the time, significantly higher than previous rates of 7.5--25.5\%, suggesting that \method\ generates highly realistic, text-accurate synthetic tumors.
\textit{\textbf{Second}, \ul{Radiomics Pattern Analysis} to analyze the diversity of generated tumor appearance.} 
We compute texture-wise Radiomics features of synthetic tumors conditioned on different random noise. \method\ exhibited much higher variance than prior arts (e.g., 1.03 for \method\ vs.~0.93 for DiffTumor~\cite{chen2024towards}; \tableautorefname~\ref{tab:variance_comparison}). This indicates that \method\ generates diverse, text-aligned, realistic tumors, explaining the robust performance of AI trained on them.
These promising results are attributable to the following novel design of \method: 

\begin{enumerate}

    \item \textbf{Text-Driven 3D Diffusion Models.} By conditioning the model on descriptive text mined from radiology reports, we achieved precise control over generated tumor characteristics such as texture, margins, and pathology type. This approach leads to more diverse and interpretable synthetic tumors (\tableautorefname~\ref{tab:variance_comparison}), thanks to the variability in text descriptions compared to random noise alone.

    \item \textbf{Text Extraction and Generation.} 
    We augmented the descriptive text by employing GPT-4o \cite{achiam2023gpt} to extract keywords and generate detailed medical reports that capture tumor characteristics such as texture, margins, and pathology types. The generated reports were automatically validated using a suite of large language models to ensure semantic consistency with the original reports. This technique is crucial when text-image pairs are scarce for training Diffusion Models.

    \item \textbf{Text-Driven Contrastive Learning.} To address the scarcity of image-report pairs (only 141 pairs used in this study), we introduced contrastive learning \cite{chen2020simple,xiao2022delving} that leverages a large corpus of 34,035 radiology reports. Positive pairs are different CT scans conditioned on the same text description; negative pairs are formed by conditioning the same CT scan on different text descriptions. We compute the contrastive loss on features extracted from the generated tumor regions, enabling the model to learn strong relationships between textual descriptions and visual tumor features.

    \item \textbf{Targeted Data Augmentation.} We analyze the failure cases (e.g., false positives and false negatives) of current state-of-the-art tumor segmentation models. Vision-language models, such as GPT-4o \cite{achiam2023gpt}, can generate detailed textual descriptions of the tumor characteristics of these failure cases, which can be used to guide diffusion models to generate a number of synthetic tumors specifically designed to enhance the segmentation models. Incorporating this targeted data augmentation led to notable improvements in tumor detection, segmentation, and classification performance.
    
\end{enumerate}

\section{Related Work}\label{sec:related_work}

Tumor synthesis has emerged as a critical research focus across various medical imaging modalities, including colonoscopy videos~\cite{shin2018abnormal}, MRI~\cite{billot2023synthseg}, CT~\cite{han2019synthesizing,lyu2022pseudo,yao2021label}, and endoscopic images~\cite{du2023boosting, wei2024sam, wei2024sam2}. While early methods~\cite{lyu2022pseudo,yao2021label,han2019synthesizing,jin2021free,wang2022anomaly,wyatt2022anoddpm,hu2023label} relied on low-level image processing techniques, their limited realism often led to noisy synthetic data that degraded model performance. To overcome these limitations, condition-guided synthesis has gained traction by enabling precise tumor localization and morphology control, facilitating data augmentation for improved detection and segmentation \cite{yang2024freemask, li2020tumor, zhang2019lung, macdm2021synthetic, wu2024freetumoradvancetumorsegmentation}. Building on this foundation, recent approaches, including conditional diffusion models and annotation-free frameworks, have further broadened its applications, addressing data scarcity across diverse medical imaging tasks \cite{hu2023label,lai2024pixel, chen2024towards}, which have been selected as baselines for this paper. However, these methods are conditioned only on shape masks, lacking controls over specific tumor characteristics (e.g., texture, heterogeneity, boundaries, and pathology).

Text-driven synthesis has emerged as a transformative tool in medical imaging, enabling the generation of diverse medical images, such as 3D scans, chest X-rays, and retinal images, based on descriptive text. This approach has significantly advanced tasks like multi-abnormality classification and rare condition research, while also improving data curation efficiency through automated labeling and synthetic data generation~\cite{hamamci2023generatect, chen2021text2image, park2020retinal,du2024boosting}. Additionally, applications in privacy-preserving analytics and digital twin technology further highlight its potential~\cite{bassi2024labelcriticdesigndata, giuffre2023harnessing}. However, existing methods primarily focus on whole-CT-level synthesis, limiting their utility for tumor-specific tasks. To address this, we develop a novel text-driven tumor synthesis framework termed \method\ that enables the precise generation of tumors based on described characteristics.

\section{\method}\label{sec:Method}

\begin{figure*}[ht]
    \centering
    \includegraphics[width=\linewidth]{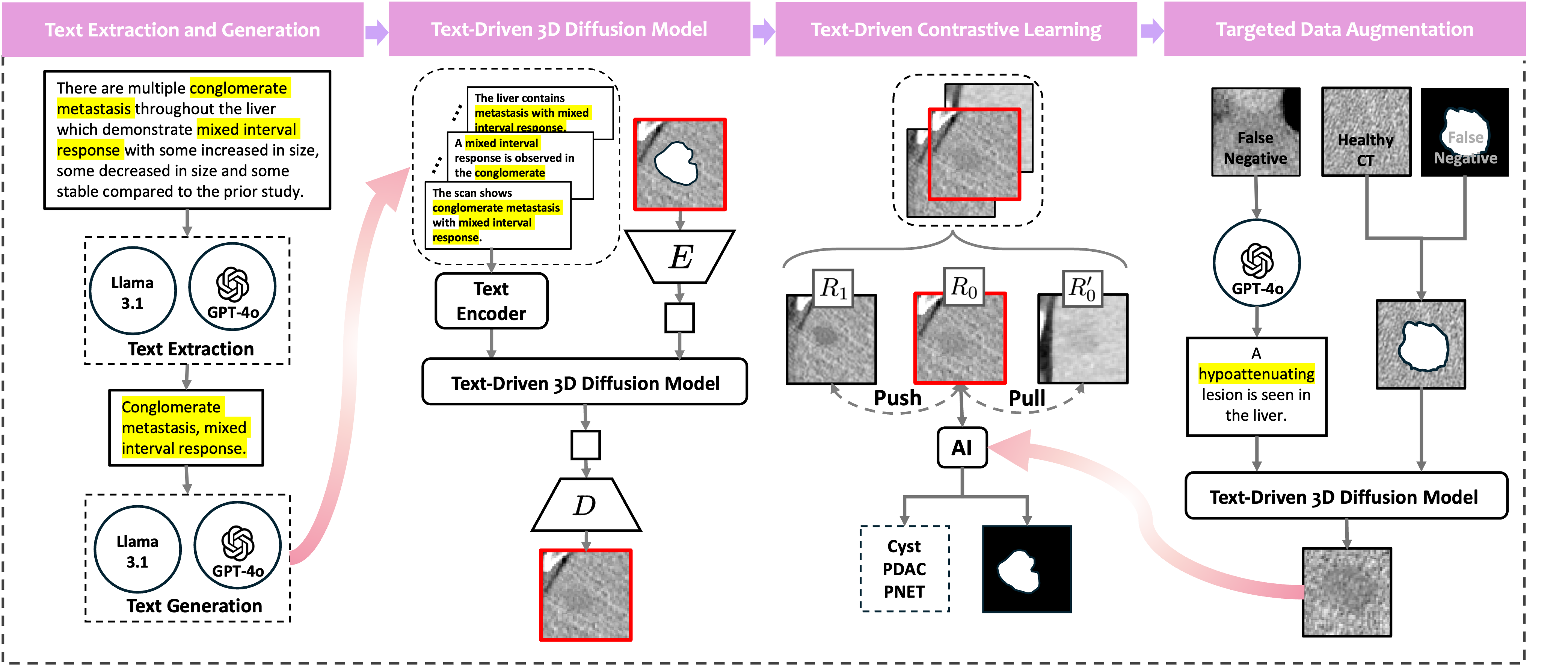}
    \caption{
    \textbf{Overview of the \method\ Framework.} The framework consists of four steps: (1) Given a radiology report, we first perform text extraction and generation to obtain descriptive phrases (e.g., \textit{conglomerate metastasis with mixed interval response}). These phrases are encoded by a text encoder (implemented via CLIP) to produce language representations guiding tumor synthesis. (2) Based on textual information and latent CT features, we train a Text-Driven 3D Diffusion Model with Encoder ($E$) and Decoder ($D$) to generate high-fidelity synthetic tumors consistent with the report descriptions. (3) Contrastive learning operations (Push vs. Pull) ensure that reports with consistent descriptive words ($R_0$ vs. $R_{0}^{\prime}$) generate similar tumors from different CTs, while distinct reports ($R_0$ vs. $R_{1}$) yield differentiable tumor features. (4) To enhance AI performance in detection, segmentation, and classification tasks, we extract descriptive texts from false positive samples to generate similar tumor examples, thereby improving the model's recognition of complex lesions.
    }
    \label{fig:model}
\end{figure*}

This paper introduces a novel framework called \method, which synthesizes tumors with realistic textures and margins using CT scans and radiology reports. \figureautorefname~\ref{fig:model} depicts the framework, consisting of (1) text extraction and generation in \S\ref{sub:text_extraction}, (2) Text-Driven 3D Diffusion Model in \S\ref{sub:LDM}, (3) Text-Driven Contrastive Learning in \S\ref{sub:contrastive}, and (4) Segmentation Model enhanced by targeted data augmentation in \S\ref{sub:targeted_data_augmentation}. In the following sections, we first introduce each component in our framework, followed by a summary of three unique properties of \method. 

\subsection{Text Extraction and Generation}\label{sub:text_extraction}
Controlling tumor synthesis through textual descriptions faces challenges due to noisy, fragmented, and inconsistent information in human-made radiology reports (see examples in \appendixautorefname~\ref{sec:supp_text_extraction}). To address these issues, we implemented a two-stage data preprocessing approach involving data cleaning and augmentation.

\smallskip\noindent\textbf{\textit{Text Extraction:}}
We employed GPT-4o to extract key tumor characteristics, focusing on features like texture and margins. Using prompts such as `\textit{Extract detailed texture and margin characteristics from the radiology report, central to the tumor field,}' we generated a cleaned descriptive output  $D_i$. To ensure that these extracted descriptions accurately reflected the original reports, we used Llama 3.1 to compute the cosine similarity between $D_i$ and original radiology report to make sure the consistency in description.

\smallskip\noindent\textbf{\textit{Text Generation:}}
For each $D_i$, we generated $N=100$ variant reports ${\mathcal{R}_{i1}, \mathcal{R}_{i2}, \ldots, \mathcal{R}_{i100}}$ by varying sentence structures while retaining core descriptive features. We prompted GPT-4o with `\textit{generate 100 reports with distinct sentence structures, ensuring that critical texture and margin information is retained accurately.}' Llama 3.1 evaluated cosine similarity to maintain alignment. This process expanded each CT image’s association from one report to 100 semantically consistent variants, forming a robust dataset $\mathcal{D} = {(x_i, \mathcal{R}_{ij})}$ for controlled tumor synthesis with enriched textual descriptions.

\subsection{Text-Driven 3D Diffusion Model} \label{sub:LDM}

We adopt Latent Diffusion Models (LDMs)~\cite{rombach2022high, lin2024stable, ddpm, yao2024addressing} for latent feature extraction from 3D CT volumes and integrate text conditioning for controlled tumor synthesis. Each 3D CT sub-volume $x \in \mathbb{R}^{H \times W \times D}$ is encoded into a lower-dimensional latent representation $z_0 = E(x)$ using a 3D VQGAN~\cite{esser2021taming} autoencoder, where $E$ is the encoder network. The decoder $D$ reconstructs the CT image from the latent representation, ensuring essential features are preserved for subsequent processing. 

To enhance radiotherapy outcomes, we follow the approach of DiffTumor \cite{chen2024towards} and choose a diffusion process with $T = 200$ time steps to generate tumors with more detailed textures. In the latent space, we define a diffusion process that progressively adds noise to the latent representation $z_0$ over discrete time steps $t = 1, \dots, T$. This noising process is predefined, and our goal is to learn the reverse denoising process using a neural network $\epsilon_\theta$. We condition the denoising model on several inputs: the healthy region latent $z_{\text{healthy}} = E((1 - m) \odot x)$ representing healthy tissue, where $m$ is the mask of tumor region; textual descriptions encoded as text embeddings $\tau_{\theta}(\mathcal{R}_i)$ derived from augmented radiology reports $\mathcal{R}_i = \{ R_{i1}, R_{i2}, \dots, R_{i100} \}$ using CLIP~\cite{clip}, where we fixed the length of augmented radiology reports by setting $|\mathcal{R}_i| = 100$, ensuring that each augmented radiology report $\mathcal{R}_i$ consists of exactly 100 segments to produce consistent text embeddings $\tau_{\theta}(\mathcal{R}_i)$; a binary tumor mask $m$ highlighting the synthesis region, and the current diffusion time step $t$, incorporated via positional encoding, are used as conditioning inputs. The denoising model $\epsilon_\theta$ predicts the noise $\hat{\epsilon}$ from the noisy latent $z_t$ at each time step $t$, using these conditioning inputs:

\begin{equation}
\hat{\epsilon} = \epsilon_\theta(z_t, t, z_{\text{healthy}}, \tau_{\theta}(\mathcal{R}_i), m).
\label{eq:predict_noise}
\end{equation}

Then, the original latent $\hat{z}_0$ is estimated based on the predicted noise $\hat{\epsilon}$:
\begin{equation}
\hat{z}_0 = \frac{1}{\sqrt{\bar{\alpha}_t}} \left( z_t - \sqrt{1 - \bar{\alpha}_t} \hat{\epsilon} \right),
\end{equation}
where \(\bar{\alpha}_t = \prod_{s=1}^t \alpha_s\) represents the cumulative noise attenuation factor in the diffusion process, with \(\alpha_s\) being the noise attenuation coefficient at each time step \(s\).

The denoising network \(\epsilon_\theta\) is a time-conditional 3D U-Net that processes the noisy latent \(z_t\) along with conditioning inputs. It integrates the healthy region latent \(z_{\text{healthy}}\) and text embeddings \(\tau_{\theta}(\mathcal{R}_i)\) through concatenation and attention mechanisms, utilizing spatial and semantic information. Spatial attention focuses on the tumor region specified by the mask \(m\), while semantic attention incorporates descriptive tumor characteristics from the text embeddings.

During inference, we start with a noisy latent representation and iteratively apply the denoising model $\epsilon_\theta$ over $T=200$ time steps, incorporating the conditioning information to guide the synthesis:
\begin{equation}
z_{t-1} = \text{Denoise}(z_t, t, z_{\text{healthy}}, \tau_{\theta}(\mathcal{R}_i), m).
\end{equation}

After $200$ steps, we obtain the reconstructed latent $\hat{z}_0$, which is then decoded using $D$ to produce the synthesized CT image with the tumor exhibiting the specified characteristics. By conditioning the latent diffusion model on healthy tissue, textual descriptions, tumor mask, and time step, and by choosing an appropriate number of diffusion steps to enhance texture details, we enable precise control over tumor synthesis in 3D CT images. This approach leverages the strengths of latent diffusion modeling and aligns with the methodology in Rombach~\etal~\cite{rombach2022high}.

\subsection{Text-Driven Contrastive Learning}\label{sub:contrastive}

\begin{figure*}[ht]
    \centering
    \includegraphics[width=\linewidth]{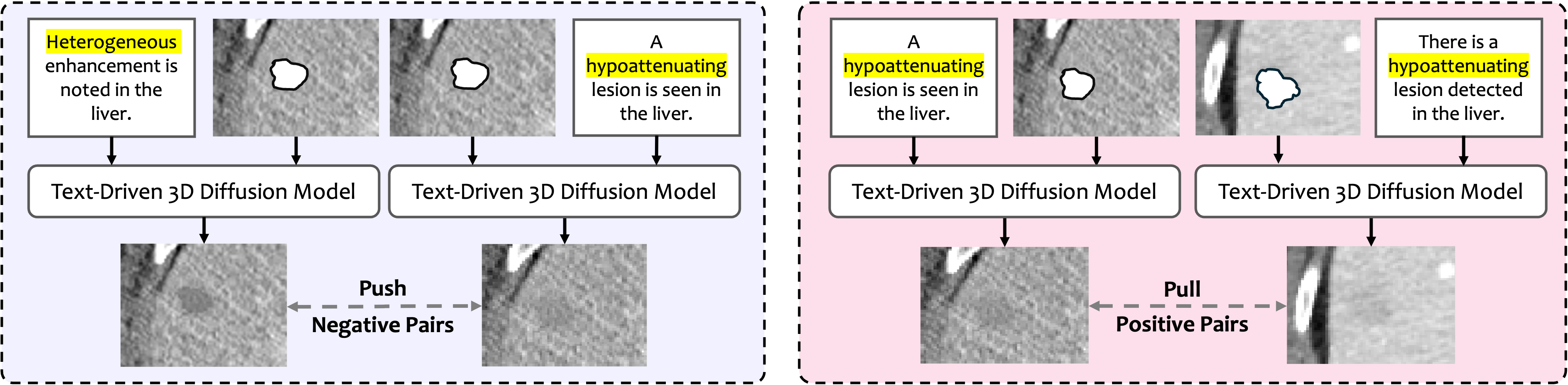}
    \caption{\textbf{Text-Driven Contrastive Learning.} We illustrate the contrastive learning approach in the diffusion model for tumor synthesis control. The negative pair shows that different descriptive words (e.g., `\textit{hypoattenuating}' vs. `\textit{heterogeneous}') applied to the same CT scan generate distinct tumors, enforcing that different descriptions lead to distinguishable features. The positive pair shows the consistent descriptive words (e.g., `\textit{hypoattenuating} vs. \textit{hypoattenuating}') applied to two different CT scans, resulting in similar tumor features, thus ensuring consistency for identical descriptions across varying CT contexts. This strategy aligns the textual descriptions with tumor synthesis, promoting both distinctiveness and consistency.
    }
    \label{fig:contrastiveloss}
\end{figure*}

We propose a structured approach for generating text-conditioned tumors using contrastive learning to improve both consistency and diversity in tumor synthesis. Each tumor generation process involves a base image \( x_0 \), paired with an initial descriptive report \( \mathcal{R}_0 \) and a corresponding mask \( m_0 \), producing the anchor tumor \( T(\mathcal{R}_0, x_0, m_0) \).
Three critical operations are incorporated to further refine and optimize the tumor generation process:


\smallskip\noindent\textbf{\textit{Push Operation:}} To encourage greater diversity, we construct a negative counterpart of the anchor tumor \( T(\mathcal{R}_0, x_0, m_0) \) by generating another tumor, \( T(\mathcal{R}_1, x_0, m_0) \), conditioned on the same CT scan and mask but paired with a distinct descriptive report \( \mathcal{R}_1 \). Here, \( R_0 \) and \( \mathcal{R}_1 \) differ in their descriptive content (e.g., highlighting varying density, margin sharpness, or boundary attributes). By increasing the feature distance between \( T(\mathcal{R}_1, x_0, m_0) \) and \( T(\mathcal{R}_0, x_0, m_0) \), \method\ effectively pushes the visual tumor features associated with different reports apart, ensuring that tumors generated from distinct textual descriptions exhibit unique and distinguishable feature characteristics.

\smallskip\noindent\textbf{\textit{Pull Operation:}} To enhance consistency, we generate a positive counterpart of the anchor tumor 
\( T(\mathcal{R}_0, x_0, m_0) \) by using a similar descriptive report 
\( \mathcal{R}_{0}^{\prime} \) with consistent descriptive terms but a different CT image 
\( x_2 \) and corresponding mask \( m_2 \). By minimizing the feature distance between 
\( T(\mathcal{R}_{0}^{\prime}, x_2, m_2) \) and \( T(\mathcal{R}_0, x_0, m_0) \), 
\method\ effectively brings the visual tumor features associated with the same 
textual description closer together. 

\smallskip\noindent\textbf{\textit{Total Loss Function:}}
The contrastive learning objective controls tumor synthesis features through similarity and diversity in the feature space, expressed as \(\mathcal{L}_{\text{same}} - \mathcal{L}_{\text{different}}\). Here, \(\mathcal{L}_{\text{same}}\) minimizes feature distances for tumors generated with the same descriptive text, while \(\mathcal{L}_{\text{different}}\) maximizes feature distances for tumors generated with distinct descriptive texts, achieving a balance between consistency and diversity in the generation process.

For the latent diffusion model process, our training loss $\mathcal{L}_{\text{ldm}}$ minimizes the difference between the predicted noise $\hat{\epsilon}$ (\cref{eq:predict_noise}) and the true noise $\epsilon$, encouraging the model to accurately predict the noise added during the noising process while adhering to the conditioning information:
\begin{equation}
\mathcal{L}_{\text{ldm}} = \mathbb{E}_{z_0, \epsilon \sim \mathcal{N}(0, 1), t} \left[ \| \epsilon - \hat{\epsilon} \|_2^2 \right].
\end{equation}

For the contrastive learning part, \( \mathcal{L}_{\text{same}} \) and \( \mathcal{L}_{\text{different}} \) ensures that tumors generated with different textual descriptions are distinguishable by maximizing the feature distance between them. This promotes diversity and specificity in the generated tumors corresponding to different text inputs.

\subsection{Targeted Data Augmentation} \label{sub:targeted_data_augmentation}

The targeted data augmentation approach (\appendixautorefname~\ref{sec:supp_targeted}) that leverages false positive (FP) tumors is introduced to enhance tumor detection and segmentation models. 

Firstly, FP tumors $\mathcal{F} = \{ (x_i, m_i) \mid i=1,\dots,n \} $ are selected from previous methods based on detection errors. Here, $x_i$ denotes a 3D CT sub-volume representing a tumor region, $m_i$ is the corresponding tumor mask indicating the tumor’s spatial extent, and $n$ is the total number of FP tumors identified. For each tumor pair $(x_i, m_i)$, we leverage GPT-4o to generate a descriptive report $\mathcal{R}_i$ that accurately reflects the tumor’s visual and semantic characteristics, including texture, shape, and edge definitions. These textual descriptions are produced through few-shot learning on a reference dataset containing frequently occurring medical terms, ensuring that the generated text comprehensively captures both the radiological and clinical features of the tumors, facilitating precise text-image alignment.

Following this, we combine a healthy CT volume $x_{\text{healthy}}$, the description $\mathcal{R}_i$, and the tumor mask $m_i$ as conditioning inputs to the Text-Driven 3D Diffusion Mode described in \S\ref{sub:LDM}. Within the latent space, noise is progressively removed while the textual and spatial constraints guide the synthesis. The text $\mathcal{R}_i$ and mask $m_i$ jointly localize and define the tumor’s features, while $x_{\text{healthy}}$ provides a baseline of normal tissue. By iteratively refining the latent representation, the model synthesizes a new CT volume enriched with challenging tumor instances that resemble real-world FP cases. The enriched training dataset helps the model better handle complex and underrepresented tumor cases, enhancing its clinical performance.

\section{Experiment and Result}\label{sec:Experiments}

\subsection{Dataset and Evaluation Metrics}\label{sub:Datasets}

\noindent\textbf{\textit{Tumor Synthesis:}} The training dataset includes 173 CT scans: 98 liver, 31 pancreas, and 78 kidney scans, each with uncertain or lesion regions, with ground truth from radiology reports, \appendixautorefname~\ref{sec:supp_data}, which presents a subset of 14 ground truth cases. True positive cases from DiffTumor~\cite{chen2024towards} were selected, consisting of 66 liver, 15 pancreas, and 60 kidney scans, totally 141 CT scans. A large corpus of 34,035 radiology reports are leveraged in Text-Driven Contrastive Learning. For evaluation, we use radiologist error rates, which measure the percentage of incorrect judgments in distinguishing real tumors from synthetic ones, to evaluate synthetic tumor realism, with higher rates indicating greater realism of synthetic tumors.

\smallskip\noindent\textbf{\textit{Tumor Segmentation:}}
We used LiTS~\cite{bilic2023liver} (131 CTs) for liver, MSD-Pancreas~\cite{antonelli2022medical} (281 CTs) for pancreas, and KiTS~\cite{heller2020international} (300 CTs) for kidney to train and test our segmentation models with a 5-fold cross-validation strategy. For healthy data, we selected CTs from the AbdomenAtlas~\cite{li2024abdomenatlas,qu2023annotating} for the liver, kidney, and pancreas, respectively, to synthesize tumors and train the segmentation model. For evaluation, we measured detection Sensitivity across small ($d < 20\,\text{mm}$), medium ($20 \leq d < 50\,\text{mm}$), and large ($d \geq 50\,\text{mm}$) tumor sizes, and assessed segmentation quality using the Dice Similarity Coefficient (DSC) and Normalized Surface Distance (NSD).

\smallskip\noindent\textbf{\textit{Tumor Classification:}} A proprietary dataset includes 5,119 CT volumes categorized into normal cases and cases with pancreatic ductal adenocarcinoma (PDAC), cysts,and pancreatic neuroendocrine tumors (PNETs) \cite{xia2022felix,chu2019utility}. The dataset is splitted into 3159 training set and 1960 test set. We constructed a small balanced dataset from the whole training set, consisting of 20 PDACs, 20 PNETs, 20 Cysts and 60 healthy. We fine-tuning \method\ on the 60 CT scans with pancreatic tumors in the training set, ensuring robust representation across all tumor categories.
For evaluation, patient-level evaluation focuses on coarse-grained detection, requiring the model to identify patients with malignant tumors in CT scans. We report Sensitivity (Sen), Specificity (Spe), and positive predictive value (PPV); tumor-level evaluation emphasizes precise localization, requiring the model to accurately locate tumors. Reported metrics include Sen, DSC, and NSD. The evaluation criteria are consistent for both benign and malignant tumors.

\begin{table}[t]
\centering
\scriptsize
    \begin{tabular}{p{0.22\linewidth}p{0.22\linewidth}P{0.1\linewidth}P{0.1\linewidth}P{0.1\linewidth}}
    \toprule
    tumor synthesis & diameter (mm) & liver & pancreas & kidney \\
    \midrule
    \multirow{3}{*}{DiffTumor \cite{chen2024towards}}
    & $d<20$ & 20.0 & 30.0& 25.5\\
    & $20\leq d<50$ & 25.5& 25.0 & 22.5 \\
    & $d \geq 50$ & 7.5 & 25.0 & 20.0\\
    \midrule
    \multirow{3}{*}{\method}
    & $d<20$ & 32.5 & 40.0& 32.5 \\
    & $20\leq d<50$ & 37.5 & 40.0 & 40.0\\
    & $d \geq 50$ & 22.5 & 45.0& 45.0 \\
    \bottomrule
    \end{tabular}
\caption{\textbf{Text-Driven Visual Turing Test.} Radiologist error rates (\%) in distinguishing real tumors from synthetic ones were evaluated for DiffTumor~\cite{chen2024towards} and the proposed \method\ across liver, pancreas, and kidney, considering tumor sizes (small: $(d < 20 mm)$, medium: $(20 \leq d <50 mm)$, large: $(d \geq 50 mm)$. Each category included 60 CT scans: 20 real tumors, 20 synthetic tumors from DiffTumor, and 20 from \method. While DiffTumor used only tumor masks, \method\ incorporated both the mask and radiology report details for enhanced realism. Higher error rates for \method\ indicate its synthetic tumors were harder for radiologists to distinguish from real ones, confirming its superior realism.}

\label{tab:visual_turing_test}
\end{table}

\subsection{Text-Driven Visual Turing Test}\label{sub:turing}
In this Visual Turing Test, radiologists evaluated a set of 540 CT scans to determine their ability to distinguish real tumors from synthetic ones. The scans were categorized based on the organ type—liver, pancreas, and kidney—and divided into three tumor size ranges: small ($d<20$ mm), medium ($20\leq d<50$ mm), and large ($d \geq 50$ mm). For each category, 60 CT scans were assessed, comprising 20 real tumors, 20 synthetic tumors generated by DiffTumor, and 20 synthetic tumors generated by \method.

Each synthetic tumor was generated based on the mask of a real tumor. However, DiffTumor relied solely on the tumor's mask, whereas \method\ incorporated both the mask and detailed information from the tumor's corresponding radiology report, aiming to improve anatomical and textural accuracy.

Radiologists were tasked with distinguishing real tumors from synthetic ones, and error rates were calculated based on instances where real tumors were misclassified as synthetic or vice versa. The higher error rates for \method\ indicate that radiologist found it more challenging to differentiate real tumors from \method-generated synthetic tumors compared to DiffTumor, reflecting a higher level of realism in \method's synthetic outputs. These findings underscore the potential of \method\ to create highly realistic synthetic tumors, potentially enhancing medical imaging data for research and training purposes.

\subsection{Tumor Detection \& Segmentation}\label{sub:ablation}
\begin{figure}[t]
  \centering
\includegraphics[width=\linewidth]{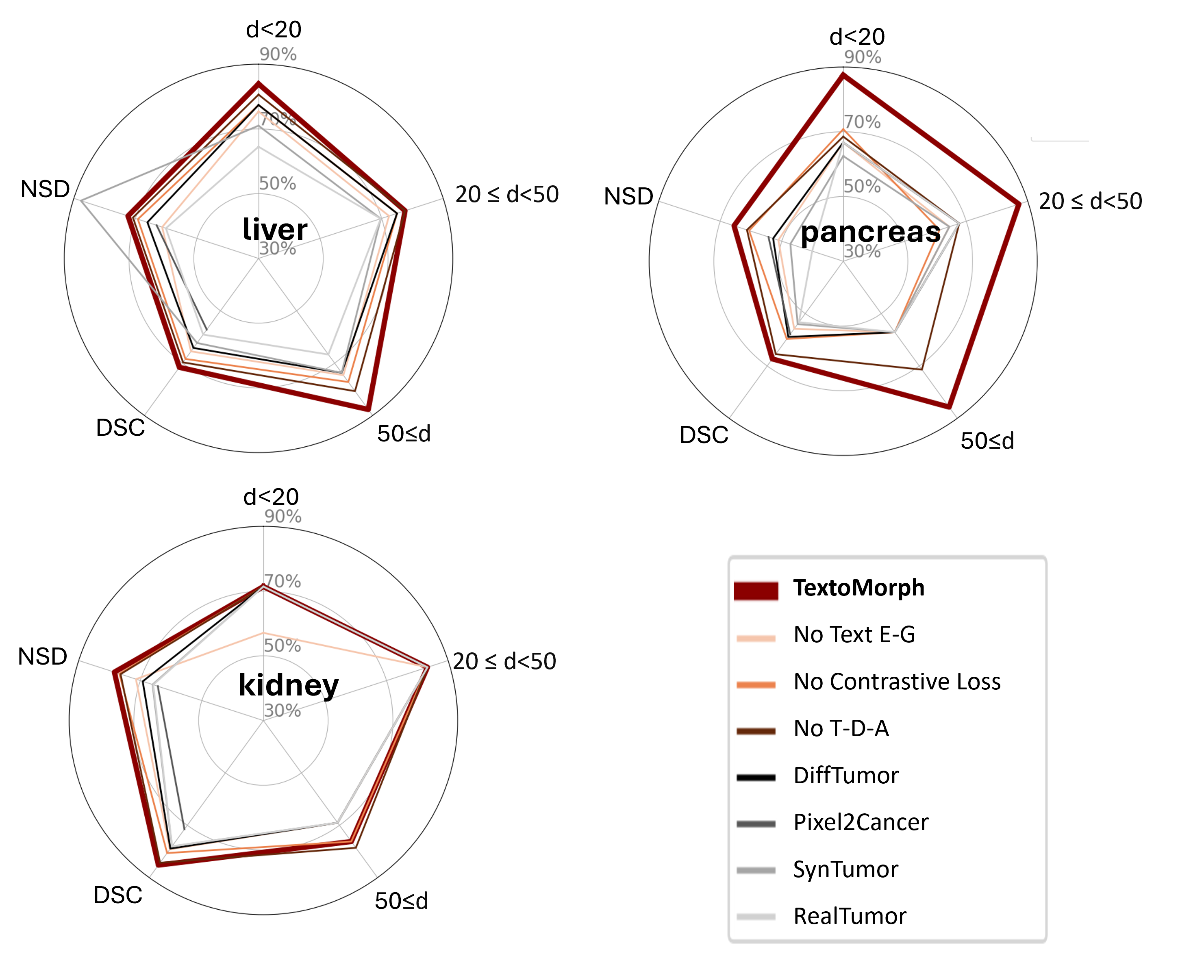}
    \caption{\textbf{Tumor Detection and Segmentation.}
    Comparison of the performance of different tumor generation models in a radial plot with an outer ring value of 90. The models include \method\ (full) and its variants excluding Text Extraction and Generation (No Text E-G), Text-Driven Contrastive Learning (No Contrastive Loss), and Targeted Data Augmentation (No T-D-A), along with DiffTumor and RealTumor. Performance metrics include sensitivity for small ($d < 20\,\mathrm{mm}$), medium ($20 \leq d < 50\,\mathrm{mm}$), and large ($d \geq 50\,\mathrm{mm}$) tumors, Dice Similarity Coefficient (DSC), and Normalized Surface Distance (NSD). Each configuration uses distinct colors or line styles to highlight the impact of individual components. See \appendixautorefname~\ref{sec:supp_Ablation_overall} for tabular results.
    }
    \label{fig:result}
\end{figure}

\smallskip\noindent\textbf{\textit{Text Extraction and Generation (\S\ref{sub:text_extraction}):}}
To evaluate the effect of Text Extraction and Generation, we compare \method\ with and without text augmentation. In the version without text augmentation, only discrete and complex radiology reports are used for tumor generation. Experimental results indicate that the version without text augmentation fails to significantly improve the AI's ability to segment and detect tumors. Specifically, for large tumors ($d \geq 50$ mm), the detection rate remains at 74.6\%, highlighting its limited capability in handling challenging cases.

\smallskip\noindent\textbf{\textit{Text-Driven Contrastive Learning (\S\ref{sub:contrastive}):}}
To study the impact of contrastive learning, we compare \method\ with and without the contrastive loss function. In this approach, the model maximizes similarity between tumors generated from similar radiology reports while increasing dissimilarity between those generated from distinct reports. This encourages the model to better capture subtle variations in tumor morphology, enhancing its ability to differentiate tumor types and improve segmentation accuracy for complex tumors, such as those with irregular borders or mixed densities. As demonstrated in \figureautorefname~\ref{fig:result} and \appendixautorefname~\ref{sec:supp_Ablation_overall}, applying contrastive learning increases the detection rate for large liver tumors by 3.5\% and improves the NSD by 1.6\%.

\smallskip\noindent\textbf{\textit{Targeted Data Augmentation (\S\ref{sub:targeted_data_augmentation}):}} 
To address the limitations of prior tumor detection methods, we introduce Targeted Data Augmentation by leveraging False Positive tumor as shown in \appendixautorefname~\ref{sec:supp_Targeted_Data_Visual_Examples}. These challenging examples are magnified and paired with descriptive text generated by GPT-4o based on tumor-specific terminology. This structured input, including tumor masks and healthy CT scans, serves as control conditions for tumor synthesis using a diffusion model as illustrate in \appendixautorefname~\ref{sec:supp_targeted}. Our approach improves the model's generalization, achieving a 4.7\% increase in DSC and a 9.1\% rise in sensitivity for large kidney tumors, as detailed in \appendixautorefname~\ref{sec:supp_Ablation_overall}.

\begin{figure}[t]
    \centering
    \includegraphics[width=\linewidth]{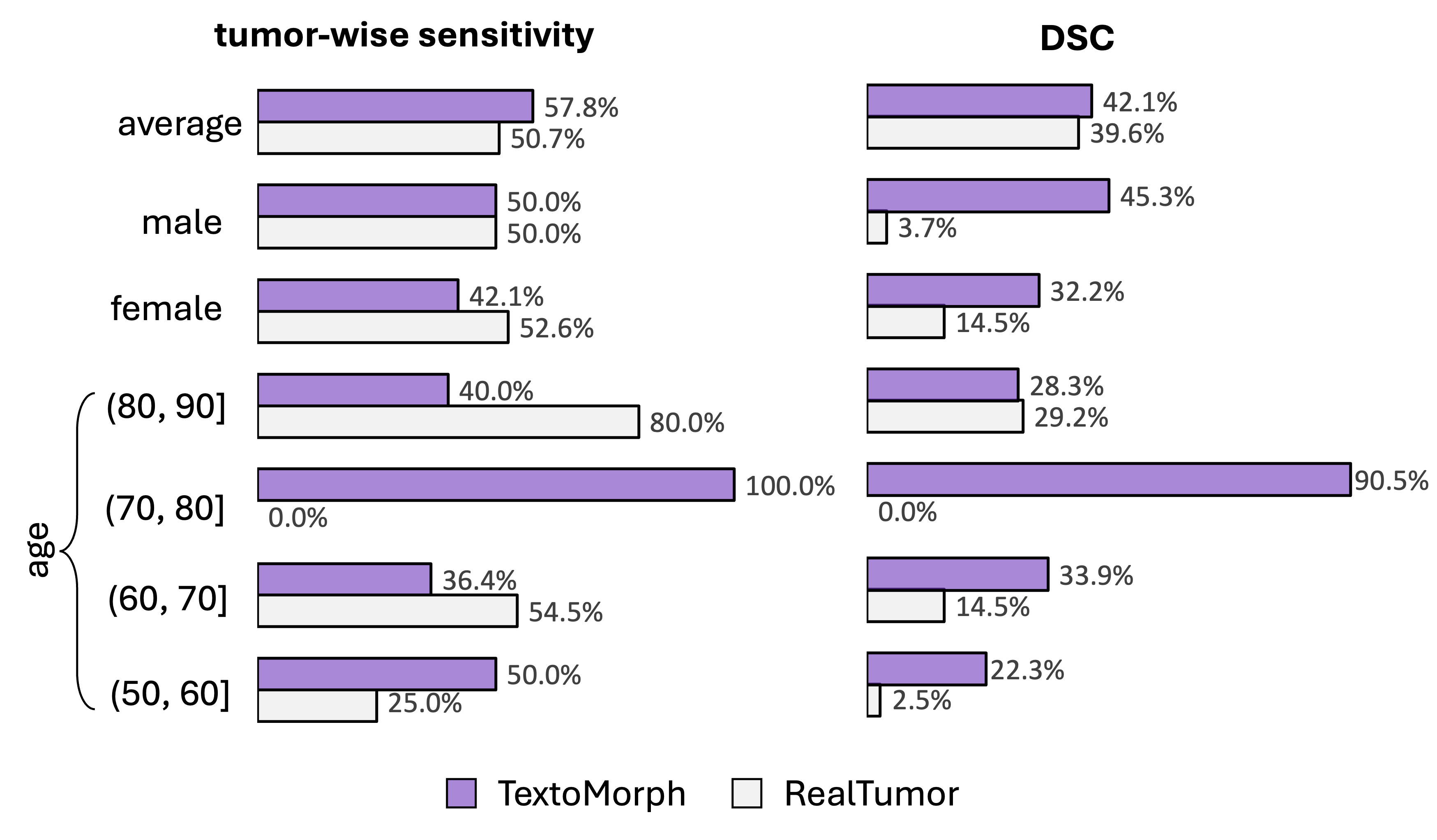}
    \caption{\textbf{Generalizable Across Different Patient Demographics.} \method\ demonstrates consistent performance improvements in detecting benign tumors (pancreatic cysts) in both tumor-wise Sensitivity (\%) and segmentation DSC (\%) across various patient groups. Results of detecting malignant tumors in the pancreas (e.g., PDAC) can be found in \appendixautorefname~\ref{sec:Generalizable}.}
    \label{fig:Generalizable}
\end{figure}

\smallskip\noindent\textbf{\textit{Generalizable to Different Demographics:}}
To evaluate the enhancement provided by \method\ in detecting and segmenting real tumors across different demographics, we used a proprietary dataset~\cite{xia2022felix,chu2019utility,kang2023label,chou2024embracing} containing pancreatic tumors (PDACs and Cysts) from patients of varying ages and genders. As shown in \figureautorefname~\ref{fig:Generalizable}, \method\ achieved 100\% sensitivity and 90.5\% DSC for the 70--80 age group, reflecting extremely strong performance for this demographic. Performance for both male and female patients also saw notable gains, further underscoring the potential of our \method\ to thoroughly bolster clinical tumor analysis across demographically diverse populations.

\subsection{Tumor Classification}\label{sub:tumor_classification}
\method\ aims to generate diverse types of tumors to improve classification accuracy and enhance model robustness. For the pancreas, the three most common tumor types are pancreatic neuroendocrine tumors (PNETs), pancreatic ductal adenocarcinoma tumors (PDACs), and cysts. To ensure comprehensive data representation across these types, descriptive sentences were generated based on each CT scan's tumor type; for instance, `\textit{a cystic lesion in the pancreas is present}' was generated for cystic tumors. For each tumor type, 20 CT scans were carefully selected, totaling 60 CT scans, to capture variations within each category and provide a solid foundation for model fine-tuning. These descriptive sentences, paired with their corresponding CT scans and tumor masks, served as inputs to fine-tune the parameters of the Text-Driven 3D diffusion model, specifically targeting these three pancreatic tumor types. 

Using the fine-tuned parameters, 40 additional synthetic instances for each type, including cysts, PNETs, and PDACs are generated and selected. These synthetic samples were then integrated with the real tumor dataset, which includes 40 scans for each tumor type and 60 for healthy cases, resulting in a balanced, diverse, and robust dataset for model training and evaluation.

As demonstrated in \tableautorefname~\ref{tab:tumor_classification}, incorporating synthetic data significantly improved tumor-level classification and segmentation metrics across various tumor types. In the RealTumor setting, the Sensitivity (Sen) for non-cyst tumors (e.g., PDAC and PNET) was 61.9\%, while cyst detection achieved 50.7\%. After initial augmentation, Sensitivity increased to 70.1\% for non-cyst tumors and 57.8\% for cysts. Further augmentation with additional synthetic data led to even greater improvements, achieving Sensitivity levels of 79.0\% for non-cyst tumors and 70.8\% for cysts, demonstrating the effectiveness of our method. Tumor-level and patient-level comparisons are presented in \tableautorefname~\ref{tab:sup_tumor_classification}.

\begin{table}[t]
\centering
\scriptsize
    \begin{tabular}
    {p{0.16\linewidth}|P{0.09\linewidth}P{0.09\linewidth}P{0.06\linewidth}|P{0.09\linewidth}P{0.09\linewidth}P{0.06\linewidth}}
    \toprule
     \multirow{3}*{method} & \multicolumn{3}{c|}{ malignant tumor } &\multicolumn{3}{c}{benign cyst} \\
    \cmidrule{2-7}
    &  Sen  & Pre  & DSC
    &  Sen  & Pre  & DSC \\
    \midrule
    RealTumor 
    &61.9\scalebox{.6}{~(304/491)} &  46.1\scalebox{.6}{~(390/846)}  & 28.1
    &50.7\scalebox{.6}{~(245/483)}  & 27.4\scalebox{.6}{~(261/953)} &  39.6 
    \\
    \method
    &70.1\scalebox{.6}{~(344/491)} &  55.2\scalebox{.6}{~(359/650)}  & 45.5 
    &57.8\scalebox{.6}{~(279/483)}  & 43.1\scalebox{.6}{~(286/663)}   & 42.1 
    \\
    \bottomrule
    \end{tabular}
\caption{\textbf{Tumor Classification Performance.} On the proprietary test dataset for RealTumor (AI trained on real data) and \method\ (AI trained on the same real data, augmented with synthetic tumors generated by \method). Tumor-level sensitivity (Sen), precision (Pre), and DSC are recorded. Results indicate that augmenting the training data with realistic synthetic tumors improves all metrics, including precision, for both non-cyst and cyst tumors in the pancreas.}
\label{tab:tumor_classification}
\end{table}

\subsection{Radiomics Pattern Analysis}\label{sub:radiomics}

To assess the diversity of the generated tumor appearances, we follow early works~\cite{zhang2023radiomics, nasief2020improving, peng2018distinguishing}, we conduct a Radiomics pattern study to evaluate the diversity of synthetic tumors generated by different methods~\cite{hu2023label,lai2024pixel, chen2024towards}. Specifically, we analyze the variance in texture-related Radiomics features~\cite{chu2019utility} to analyze how well the models capture tumor heterogeneity across different organs. 
In this experiment, we compute 102 texture-wise Radiomics features, including intensity and texture, extracted from a total of 480 synthetic tumors (120 per method, 40 per organ). For a fair comparison, these method use the same tumor masks and healthy CT scans as spatial and background constraints during generation, while, textual descriptions are used as conditions for \method\ to synthesize tumor.

Comprehensive evaluation is performed to assess each model's ability to produce diverse and heterogeneous tumor appearances. Mean variance (MV) and standard deviation (SD) of pairwise cosine similarities between Radiomics feature vectors quantify texture breadth and range: higher MV signifies greater overall diversity, while higher SD indicates a wider variety of texture patterns.

\begin{table}[t]
    \centering
    \scriptsize
    \begin{tabular}{p{0.28\linewidth}P{0.17\linewidth}P{0.17\linewidth}P{0.17\linewidth}}
    \toprule
    methods & liver & pancreas & kidney \\
    \midrule
    SynTumor~\cite{hu2023label} &1.03$\pm$0.84&0.92$\pm$0.99 &0.89$\pm$0.80 \\
    Pixel2Cancer~\cite{lai2024pixel} & 0.99$\pm$0.92&0.91$\pm$0.83&1.00$\pm$1.01 \\
    DiffTumor~\cite{chen2024towards} & 1.09$\pm$0.89& 0.95$\pm$0.90& 0.93$\pm$0.88\\
    \midrule
    \method & 1.14$\pm$1.02& 0.94$\pm$0.92& 1.03$\pm$1.02\\
    \bottomrule
    \end{tabular}
\caption{\textbf{Radiomics Pattern Analysis.} This analysis compares the mean variance (MV) $\pm$ standard deviation (SD) of texture features from synthetic tumors generated by baselines and \method\ across liver, pancreas, and kidney organs. \method\ leverages different descriptive words from medical reports as conditioning inputs. Radiomics features (e.g., intensity and texture) are extracted from 120 synthetic tumors per mathod. Pairwise cosine similarity scores between Radiomics feature vectors are used to calculate MV and SD. Higher MV and SD values for \method\ indicate its ability to produce tumors with greater heterogeneity.
}
\label{tab:variance_comparison}
\end{table}

As shown in \tableautorefname~\ref{tab:variance_comparison}, we compare the MV and SD of texture features across liver, pancreas, and kidney tumors for both \method\ and DiffTumor models. The results demonstrate that \method\ exhibited significantly higher value in Radiomics features than prior methods (1.03 for \method\ vs. 1.00 for Pixel2Cancer~\cite{lai2024pixel}) in kidney tumors and (1.14 for \method\ vs. 1.03 for SynTumor~\cite{hu2023label}) in liver tumors, underscoring its enhanced capacity to produce a more diverse and heterogeneous set of synthetic tumor textures. 
Despite the MV values for pancreas tumors being nearly identical (0.94 for \method\ vs. 0.95 for DiffTumor~\cite{chen2024towards}), \method's slightly higher SD (0.92 for \method\ vs. 0.90 for DiffTumor~\cite{chen2024towards}) suggests it generates a broader range of features, indicating greater diversity and thus better overall performance, which is crucial for achieving realistic data augmentation.

\section{Conclusion}\label{sub:conclusion}

\method\ improves AI for cancer imaging by generating realistic, diverse tumors with fine-grained control over key characteristics in CT scans, such as texture, boundaries, size, and pathology. By exploiting descriptive text from radiology reports, \method\ addresses the limitations of existing synthesis methods, enabling targeted data augmentation to create tumors that AI models often miss due to the scarcity of training CT scans with real tumors, leading to significant improvements in tumor detection, segmentation, and classification. Furthermore, this text-driven synthesis reduces reliance on scarce annotated medical datasets, offering a scalable and efficient solution to augment medical imaging data and better address critical clinical needs.

\medskip\noindent\textbf{Acknowledgments.}
This work was supported by the Lustgarten Foundation for Pancreatic Cancer Research and the Patrick J. McGovern Foundation Award.

\clearpage
{
    \small
    \bibliographystyle{ieeenat_fullname}
    \bibliography{zzhou,refs}

\begin{thebibliography}{66}
\providecommand{\natexlab}[1]{#1}
\providecommand{\url}[1]{\texttt{#1}}
\expandafter\ifx\csname urlstyle\endcsname\relax
  \providecommand{\doi}[1]{doi: #1}\else
  \providecommand{\doi}{doi: \begingroup \urlstyle{rm}\Url}\fi

\bibitem[Achiam et~al.(2023)Achiam, Adler, Agarwal, Ahmad, Akkaya, Aleman, Almeida, Altenschmidt, Altman, Anadkat, et~al.]{achiam2023gpt}
Josh Achiam, Steven Adler, Sandhini Agarwal, Lama Ahmad, Ilge Akkaya, Florencia~Leoni Aleman, Diogo Almeida, Janko Altenschmidt, Sam Altman, Shyamal Anadkat, et~al.
\newblock Gpt-4 technical report.
\newblock \emph{arXiv preprint arXiv:2303.08774}, 2023.

\bibitem[Antonelli et~al.(2022)Antonelli, Reinke, Bakas, Farahani, Kopp-Schneider, Landman, Litjens, Menze, Ronneberger, Summers, et~al.]{antonelli2022medical}
Michela Antonelli, Annika Reinke, Spyridon Bakas, Keyvan Farahani, Annette Kopp-Schneider, Bennett~A Landman, Geert Litjens, Bjoern Menze, Olaf Ronneberger, Ronald~M Summers, et~al.
\newblock The medical segmentation decathlon.
\newblock \emph{Nature communications}, 13\penalty0 (1):\penalty0 4128, 2022.

\bibitem[Basaran et~al.(2023)Basaran, Zhang, Qiao, Kainz, Matthews, and Bai]{basaran2023lesionmix}
Berke~Doga Basaran, Weitong Zhang, Mengyun Qiao, Bernhard Kainz, Paul~M Matthews, and Wenjia Bai.
\newblock Lesionmix: A lesion-level data augmentation method for medical image segmentation.
\newblock In \emph{International Conference on Medical Image Computing and Computer-Assisted Intervention}, pages 73--83. Springer, 2023.

\bibitem[Bassi et~al.(2024{\natexlab{a}})Bassi, Li, Tang, Isensee, Wang, Chen, Chou, Kirchhoff, Rokuss, Huang, Ye, He, Wald, Ulrich, Baumgartner, Roy, Maier-Hein, Jaeger, Ye, Xie, Zhang, Chen, Xia, Xing, Zhu, Sadegheih, Bozorgpour, Kumari, Azad, Merhof, Shi, Ma, Du, Bai, Huang, Zhao, Wang, Li, Gu, Dong, Yang, Mazurowski, Gupta, Wu, Zhuang, Chen, Roth, Xu, Blaschko, Decherchi, Cavalli, Yuille, and Zhou]{bassi2024touchstone}
Pedro~RAS Bassi, Wenxuan Li, Yucheng Tang, Fabian Isensee, Zifu Wang, Jieneng Chen, Yu-Cheng Chou, Yannick Kirchhoff, Maximilian Rokuss, Ziyan Huang, Jin Ye, Junjun He, Tassilo Wald, Constantin Ulrich, Michael Baumgartner, Saikat Roy, Klaus~H. Maier-Hein, Paul Jaeger, Yiwen Ye, Yutong Xie, Jianpeng Zhang, Ziyang Chen, Yong Xia, Zhaohu Xing, Lei Zhu, Yousef Sadegheih, Afshin Bozorgpour, Pratibha Kumari, Reza Azad, Dorit Merhof, Pengcheng Shi, Ting Ma, Yuxin Du, Fan Bai, Tiejun Huang, Bo Zhao, Haonan Wang, Xiaomeng Li, Hanxue Gu, Haoyu Dong, Jichen Yang, Maciej~A. Mazurowski, Saumya Gupta, Linshan Wu, Jiaxin Zhuang, Hao Chen, Holger Roth, Daguang Xu, Matthew~B. Blaschko, Sergio Decherchi, Andrea Cavalli, Alan~L. Yuille, and Zongwei Zhou.
\newblock Touchstone benchmark: Are we on the right way for evaluating ai algorithms for medical segmentation?
\newblock \emph{Conference on Neural Information Processing Systems}, 2024{\natexlab{a}}.

\bibitem[Bassi et~al.(2024{\natexlab{b}})Bassi, Wu, Li, Decherchi, Cavalli, Yuille, and Zhou]{bassi2024labelcriticdesigndata}
Pedro R. A.~S. Bassi, Qilong Wu, Wenxuan Li, Sergio Decherchi, Andrea Cavalli, Alan Yuille, and Zongwei Zhou.
\newblock Label critic: Design data before models, 2024{\natexlab{b}}.

\bibitem[Bilic et~al.(2019)Bilic, Christ, Vorontsov, Chlebus, Chen, Dou, Fu, Han, Heng, Hesser, et~al.]{bilic2019liver}
Patrick Bilic, Patrick~Ferdinand Christ, Eugene Vorontsov, Grzegorz Chlebus, Hao Chen, Qi Dou, Chi-Wing Fu, Xiao Han, Pheng-Ann Heng, J{\"u}rgen Hesser, et~al.
\newblock The liver tumor segmentation benchmark (lits).
\newblock \emph{arXiv preprint arXiv:1901.04056}, 2019.

\bibitem[Bilic et~al.(2023)Bilic, Christ, Li, Vorontsov, Ben-Cohen, Kaissis, Szeskin, Jacobs, Mamani, Chartrand, et~al.]{bilic2023liver}
Patrick Bilic, Patrick Christ, Hongwei~Bran Li, Eugene Vorontsov, Avi Ben-Cohen, Georgios Kaissis, Adi Szeskin, Colin Jacobs, Gabriel Efrain~Humpire Mamani, Gabriel Chartrand, et~al.
\newblock The liver tumor segmentation benchmark (lits).
\newblock \emph{Medical Image Analysis}, 84:\penalty0 102680, 2023.

\bibitem[Billot et~al.(2023)]{billot2023synthseg}
Benjamin Billot et~al.
\newblock Synthseg: Segmentation of brain mri scans of any contrast and resolution without retraining.
\newblock \emph{Medical Image Analy.}, 86:\penalty0 102789, 2023.

\bibitem[Cardoso et~al.(2022)Cardoso, Li, Brown, Ma, Kerfoot, Wang, Murrey, Myronenko, Zhao, Yang, et~al.]{cardoso2022monai}
M~Jorge Cardoso, Wenqi Li, Richard Brown, Nic Ma, Eric Kerfoot, Yiheng Wang, Benjamin Murrey, Andriy Myronenko, Can Zhao, Dong Yang, et~al.
\newblock Monai: An open-source framework for deep learning in healthcare.
\newblock \emph{arXiv preprint arXiv:2211.02701}, 2022.

\bibitem[Chen et~al.(2024{\natexlab{a}})Chen, Chen, Song, Xiong, Yuille, Wei, and Zhou]{chen2024towards}
Qi Chen, Xiaoxi Chen, Haorui Song, Zhiwei Xiong, Alan Yuille, Chen Wei, and Zongwei Zhou.
\newblock Towards generalizable tumor synthesis.
\newblock In \emph{IEEE/CVF Conference on Computer Vision and Pattern Recognition}, 2024{\natexlab{a}}.

\bibitem[Chen et~al.(2024{\natexlab{b}})Chen, Lai, Chen, Hu, Yuille, and Zhou]{chen2024analyzing}
Qi Chen, Yuxiang Lai, Xiaoxi Chen, Qixin Hu, Alan Yuille, and Zongwei Zhou.
\newblock Analyzing tumors by synthesis.
\newblock \emph{arXiv preprint arXiv:2409.06035}, 2024{\natexlab{b}}.

\bibitem[Chen et~al.(2020)Chen, Kornblith, Norouzi, and Hinton]{chen2020simple}
Ting Chen, Simon Kornblith, Mohammad Norouzi, and Geoffrey Hinton.
\newblock A simple framework for contrastive learning of visual representations.
\newblock In \emph{International conference on machine learning}, pages 1597--1607. PMLR, 2020.

\bibitem[Chen et~al.(2021)Chen, Li, and Zhu]{chen2021text2image}
Xiaoyu Chen, Yifan Li, and Yan Zhu.
\newblock Text2image: Synthesizing chest x-rays from radiology reports using generative adversarial networks.
\newblock In \emph{Proceedings of the IEEE Conference on Computer Vision and Pattern Recognition}, pages 10--19. IEEE, 2021.

\bibitem[Chou et~al.(2024)Chou, Zhou, and Yuille]{chou2024embracing}
Yu-Cheng Chou, Zongwei Zhou, and Alan Yuille.
\newblock Embracing massive medical data.
\newblock In \emph{International Conference on Medical Image Computing and Computer-Assisted Intervention}, pages 24--35. Springer, 2024.

\bibitem[Chu et~al.(2019)Chu, Park, Kawamoto, Fouladi, Shayesteh, Zinreich, Graves, Horton, Hruban, Yuille, et~al.]{chu2019utility}
Linda~C Chu, Seyoun Park, Satomi Kawamoto, Daniel~F Fouladi, Shahab Shayesteh, Eva~S Zinreich, Jefferson~S Graves, Karen~M Horton, Ralph~H Hruban, Alan~L Yuille, et~al.
\newblock Utility of ct radiomics features in differentiation of pancreatic ductal adenocarcinoma from normal pancreatic tissue.
\newblock \emph{American Journal of Roentgenology}, 213\penalty0 (2):\penalty0 349--357, 2019.

\bibitem[Doe et~al.(2021)Doe, Smith, and Lee]{macdm2021synthetic}
John Doe, Jane Smith, and Michael Lee.
\newblock Mac-dm: Mask-controlled diffusion models for synthetic distal tibial radiographs.
\newblock \emph{Medical Image Analysis}, 67:\penalty0 101812, 2021.

\bibitem[Du et~al.(2024)Du, Wang, Lu, Zhou, Zhang, Yuille, Li, and Zhou]{du2024boosting}
Shiyi Du, Xiaosong Wang, Yongyi Lu, Yuyin Zhou, Shaoting Zhang, Alan Yuille, Kang Li, and Zongwei Zhou.
\newblock Boosting dermatoscopic lesion segmentation via diffusion models with visual and textual prompts.
\newblock In \emph{2024 IEEE International Symposium on Biomedical Imaging (ISBI)}, pages 1--5. IEEE, 2024.

\bibitem[Du et~al.(2023)]{du2023boosting}
Shiyi Du et~al.
\newblock Boosting dermatoscopic lesion segmentation via diffusion models with visual and textual prompts.
\newblock \emph{arXiv preprint arXiv:2310.02906}, 2023.

\bibitem[Esser et~al.(2021)Esser, Rombach, and Ommer]{esser2021taming}
Patrick Esser, Robin Rombach, and Bjorn Ommer.
\newblock Taming transformers for high-resolution image synthesis.
\newblock In \emph{Proceedings of the IEEE/CVF conference on computer vision and pattern recognition}, pages 12873--12883, 2021.

\bibitem[Giuffr{\`e} et~al.(2023)Giuffr{\`e}, Romano, and Vitale]{giuffre2023harnessing}
Mauro Giuffr{\`e}, Francesco Romano, and Fabio Vitale.
\newblock Harnessing synthetic data in healthcare: Applications, challenges, and future directions.
\newblock \emph{Journal of Medical Systems}, 47\penalty0 (2):\penalty0 45, 2023.

\bibitem[Gon{\c{c}}alves et~al.(2024)Gon{\c{c}}alves, Silva, Vieira, and Vieira]{gonccalves2024abdominal}
Bernardo Gon{\c{c}}alves, Mariana Silva, Lu{\'\i}sa Vieira, and Pedro Vieira.
\newblock Abdominal mri unconditional synthesis with medical assessment.
\newblock \emph{BioMedInformatics}, 4\penalty0 (2):\penalty0 1506--1518, 2024.

\bibitem[Hamamci et~al.(2023)Hamamci, Kantarci, and Yaman]{hamamci2023generatect}
Mehmet Hamamci, Utku Kantarci, and Burak Yaman.
\newblock Generatect: Text-guided 3d medical image synthesis for multi-abnormality classification.
\newblock \emph{IEEE Journal of Biomedical and Health Informatics}, 27\penalty0 (4):\penalty0 1234--1245, 2023.

\bibitem[Han et~al.(2019)]{han2019synthesizing}
Changhee Han et~al.
\newblock Synthesizing diverse lung nodules wherever massively: 3d multi-conditional gan-based ct image augmentation for object detection.
\newblock In \emph{3DV}, pages 729--737. IEEE, 2019.

\bibitem[Heller et~al.(2020)Heller, McSweeney, Peterson, Peterson, Rickman, Stai, Tejpaul, Oestreich, Blake, Rosenberg, et~al.]{heller2020international}
Nicholas Heller, Sean McSweeney, Matthew~Thomas Peterson, Sarah Peterson, Jack Rickman, Bethany Stai, Resha Tejpaul, Makinna Oestreich, Paul Blake, Joel Rosenberg, et~al.
\newblock An international challenge to use artificial intelligence to define the state-of-the-art in kidney and kidney tumor segmentation in ct imaging., 2020.

\bibitem[Heller et~al.(2023)Heller, Isensee, Trofimova, Tejpaul, Zhao, Chen, Wang, Golts, Khapun, Shats, Shoshan, Gilboa-Solomon, George, Yang, Zhang, Zhang, Xia, Wu, Liu, Walczak, McSweeney, Vasdev, Hornung, Solaiman, Schoephoerster, Abernathy, Wu, Abdulkadir, Byun, Spriggs, Struyk, Austin, Simpson, Hagstrom, Virnig, French, Venkatesh, Chan, Moore, Jacobsen, Austin, Austin, Regmi, Papanikolopoulos, and Weight]{heller2023kits21}
Nicholas Heller, Fabian Isensee, Dasha Trofimova, Resha Tejpaul, Zhongchen Zhao, Huai Chen, Lisheng Wang, Alex Golts, Daniel Khapun, Daniel Shats, Yoel Shoshan, Flora Gilboa-Solomon, Yasmeen George, Xi Yang, Jianpeng Zhang, Jing Zhang, Yong Xia, Mengran Wu, Zhiyang Liu, Ed Walczak, Sean McSweeney, Ranveer Vasdev, Chris Hornung, Rafat Solaiman, Jamee Schoephoerster, Bailey Abernathy, David Wu, Safa Abdulkadir, Ben Byun, Justice Spriggs, Griffin Struyk, Alexandra Austin, Ben Simpson, Michael Hagstrom, Sierra Virnig, John French, Nitin Venkatesh, Sarah Chan, Keenan Moore, Anna Jacobsen, Susan Austin, Mark Austin, Subodh Regmi, Nikolaos Papanikolopoulos, and Christopher Weight.
\newblock The kits21 challenge: Automatic segmentation of kidneys, renal tumors, and renal cysts in corticomedullary-phase ct, 2023.

\bibitem[Ho et~al.(2020)Ho, Jain, and Abbeel]{ddpm}
Jonathan Ho, Ajay Jain, and Pieter Abbeel.
\newblock Denoising diffusion probabilistic models, 2020.

\bibitem[Ho et~al.(2022)Ho, Chan, Saharia, Whang, Gao, Gritsenko, Kingma, Poole, Norouzi, Fleet, et~al.]{ho2022imagen}
Jonathan Ho, William Chan, Chitwan Saharia, Jay Whang, Ruiqi Gao, Alexey Gritsenko, Diederik~P Kingma, Ben Poole, Mohammad Norouzi, David~J Fleet, et~al.
\newblock Imagen video: High definition video generation with diffusion models.
\newblock \emph{arXiv preprint arXiv:2210.02303}, 2022.

\bibitem[Hu et~al.(2022)Hu, Xiao, Chen, Sun, Chen, Yuille, and Zhou]{hu2022synthetic}
Qixin Hu, Junfei Xiao, Yixiong Chen, Shuwen Sun, Jie-Neng Chen, Alan Yuille, and Zongwei Zhou.
\newblock Synthetic tumors make ai segment tumors better.
\newblock \emph{NeurIPS Workshop on Medical Imaging meets NeurIPS}, 2022.

\bibitem[Hu et~al.(2023)Hu, Chen, Xiao, Sun, Chen, Yuille, and Zhou]{hu2023label}
Qixin Hu, Yixiong Chen, Junfei Xiao, Shuwen Sun, Jieneng Chen, Alan~L Yuille, and Zongwei Zhou.
\newblock Label-free liver tumor segmentation.
\newblock In \emph{IEEE/CVF Conference on Computer Vision and Pattern Recognition}, pages 7422--7432, 2023.

\bibitem[Jin et~al.(2021)Jin, Cui, Sun, Meng, and Su]{jin2021free}
Qiangguo Jin, Hui Cui, Changming Sun, Zhaopeng Meng, and Ran Su.
\newblock Free-form tumor synthesis in computed tomography images via richer generative adversarial network.
\newblock \emph{Knowledge-Based Systems}, 218:\penalty0 106753, 2021.

\bibitem[Kang et~al.(2023)Kang, Li, Zhu, Lu, Fishman, Yuille, and Zhou]{kang2023label}
Mintong Kang, Bowen Li, Zengle Zhu, Yongyi Lu, Elliot~K Fishman, Alan Yuille, and Zongwei Zhou.
\newblock Label-assemble: Leveraging multiple datasets with partial labels.
\newblock In \emph{IEEE International Symposium on Biomedical Imaging}, pages 1--5. IEEE, 2023.

\bibitem[Lai et~al.(2024)Lai, Chen, Wang, Yuille, and Zhou]{lai2024pixel}
Yuxiang Lai, Xiaoxi Chen, Angtian Wang, Alan Yuille, and Zongwei Zhou.
\newblock From pixel to cancer: Cellular automata in computed tomography.
\newblock \emph{arXiv preprint arXiv:2403.06459}, 2024.

\bibitem[Li et~al.(2023)Li, Chou, Sun, Qiao, Yuille, and Zhou]{li2023early}
Bowen Li, Yu-Cheng Chou, Shuwen Sun, Hualin Qiao, Alan Yuille, and Zongwei Zhou.
\newblock Early detection and localization of pancreatic cancer by label-free tumor synthesis.
\newblock \emph{MICCAI Workshop on Big Task Small Data, 1001-AI}, 2023.

\bibitem[Li et~al.(2020)Li, Fan, and Wu]{li2020tumor}
Haochen Li, Yibo Fan, and Jian Wu.
\newblock Tumor synthesis with adversarial networks for augmenting data in medical imaging.
\newblock \emph{IEEE Transactions on Medical Imaging}, 39\penalty0 (7):\penalty0 2380--2390, 2020.

\bibitem[Li et~al.(2024)Li, Qu, Chen, Bassi, Shi, Lai, Yu, Xue, Chen, Lin, et~al.]{li2024abdomenatlas}
Wenxuan Li, Chongyu Qu, Xiaoxi Chen, Pedro~RAS Bassi, Yijia Shi, Yuxiang Lai, Qian Yu, Huimin Xue, Yixiong Chen, Xiaorui Lin, et~al.
\newblock Abdomenatlas: A large-scale, detailed-annotated, \& multi-center dataset for efficient transfer learning and open algorithmic benchmarking.
\newblock \emph{Medical Image Analysis}, page 103285, 2024.

\bibitem[Lin et~al.(2024)Lin, Chen, Yan, Yu, and Zheng]{lin2024stable}
Tianyu Lin, Zhiguang Chen, Zhonghao Yan, Weijiang Yu, and Fudan Zheng.
\newblock Stable diffusion segmentation for biomedical images with single-step reverse process.
\newblock In \emph{Medical Image Computing and Computer Assisted Intervention -- MICCAI 2024}, pages 656--666, Cham, 2024. Springer Nature Switzerland.

\bibitem[Liu et~al.(2023)Liu, Zhang, Chen, Xiao, Lu, A~Landman, Yuan, Yuille, Tang, and Zhou]{liu2023clip}
Jie Liu, Yixiao Zhang, Jie-Neng Chen, Junfei Xiao, Yongyi Lu, Bennett A~Landman, Yixuan Yuan, Alan Yuille, Yucheng Tang, and Zongwei Zhou.
\newblock Clip-driven universal model for organ segmentation and tumor detection.
\newblock In \emph{Proceedings of the IEEE/CVF International Conference on Computer Vision}, pages 21152--21164, 2023.

\bibitem[Liu et~al.(2024)Liu, Zhang, Wang, Yavuz, Chen, Yuan, Li, Yang, Yuille, Tang, et~al.]{liu2024universal}
Jie Liu, Yixiao Zhang, Kang Wang, Mehmet~Can Yavuz, Xiaoxi Chen, Yixuan Yuan, Haoliang Li, Yang Yang, Alan Yuille, Yucheng Tang, et~al.
\newblock Universal and extensible language-vision models for organ segmentation and tumor detection from abdominal computed tomography.
\newblock \emph{Medical Image Analysis}, page 103226, 2024.

\bibitem[Lyu et~al.(2022)]{lyu2022pseudo}
Fei Lyu et~al.
\newblock Pseudo-label guided image synthesis for semi-supervised covid-19 pneumonia infection segmentation.
\newblock \emph{IEEE Trans. Medical Imag.}, 42\penalty0 (3):\penalty0 797--809, 2022.

\bibitem[Nasief et~al.(2020)Nasief, Hall, Zheng, Tsai, and Wang]{nasief2020improving}
Haidy Nasief, William Hall, Cheng Zheng, Susan Tsai, and Liang Wang.
\newblock Improving treatment response prediction for chemoradiation therapy of pancreatic cancer using a combination of delta-radiomics and the clinical biomarker ca19-9.
\newblock \emph{Frontiers in Oncology}, 10:\penalty0 203, 2020.

\bibitem[Nichol et~al.(2021)Nichol, Dhariwal, Ramesh, Shyam, Mishkin, McGrew, Sutskever, and Chen]{nichol2021glide}
Alex Nichol, Prafulla Dhariwal, Aditya Ramesh, Pranav Shyam, Pamela Mishkin, Bob McGrew, Ilya Sutskever, and Mark Chen.
\newblock Glide: Towards photorealistic image generation and editing with text-guided diffusion models.
\newblock \emph{arXiv preprint arXiv:2112.10741}, 2021.

\bibitem[Niemeijer et~al.(2024)Niemeijer, Ehrhardt, Uzunova, and Handels]{niemeijer2024tsynd}
Joshua Niemeijer, Jan Ehrhardt, Hristina Uzunova, and Heinz Handels.
\newblock Tsynd: Targeted synthetic data generation for enhanced medical image classification: Leveraging epistemic uncertainty to improve model performance.
\newblock In \emph{International Workshop on Simulation and Synthesis in Medical Imaging}, pages 69--78. Springer, 2024.

\bibitem[Park et~al.(2020)Park, Lee, and Kang]{park2020retinal}
Sungjun Park, Hoyeon Lee, and Minyoung Kang.
\newblock Generating retinal images from text descriptions for ophthalmology applications.
\newblock \emph{Medical Image Analysis}, 64:\penalty0 101741, 2020.

\bibitem[Peng et~al.(2018)Peng, Parekh, Huang, Lin, Sheikh, Baker, Kirschbaum, Silvestri, Son, Robinson, et~al.]{peng2018distinguishing}
Lily Peng, Vishwa Parekh, Pei Huang, Dandan~D Lin, Khurram Sheikh, Brad Baker, Thomas Kirschbaum, Frank Silvestri, Joon Son, Andrew Robinson, et~al.
\newblock Distinguishing true progression from radionecrosis after stereotactic radiation therapy for brain metastases with machine learning and radiomics.
\newblock \emph{International Journal of Radiation Oncology* Biology* Physics}, 102\penalty0 (4):\penalty0 1236--1243, 2018.

\bibitem[Qu et~al.(2023)Qu, Zhang, Qiao, Liu, Tang, Yuille, and Zhou]{qu2023annotating}
Chongyu Qu, Tiezheng Zhang, Hualin Qiao, Jie Liu, Yucheng Tang, Alan Yuille, and Zongwei Zhou.
\newblock Abdomenatlas-8k: Annotating 8,000 abdominal ct volumes for multi-organ segmentation in three weeks.
\newblock In \emph{Conference on Neural Information Processing Systems}, 2023.

\bibitem[Radford et~al.(2021)Radford, Kim, Hallacy, Ramesh, Goh, Agarwal, Sastry, Askell, Mishkin, Clark, Krueger, and Sutskever]{clip}
Alec Radford, Jong~Wook Kim, Chris Hallacy, Aditya Ramesh, Gabriel Goh, Sandhini Agarwal, Girish Sastry, Amanda Askell, Pamela Mishkin, Jack Clark, Gretchen Krueger, and Ilya Sutskever.
\newblock Learning transferable visual models from natural language supervision, 2021.

\bibitem[Ramesh et~al.(2021)Ramesh, Pavlov, Goh, Gray, Voss, Radford, Chen, and Sutskever]{ramesh2021zero}
Aditya Ramesh, Mikhail Pavlov, Gabriel Goh, Scott Gray, Chelsea Voss, Alec Radford, Mark Chen, and Ilya Sutskever.
\newblock Zero-shot text-to-image generation.
\newblock In \emph{International conference on machine learning}, pages 8821--8831. Pmlr, 2021.

\bibitem[Rombach et~al.(2022)Rombach, Blattmann, Lorenz, Esser, and Ommer]{rombach2022high}
Robin Rombach, Andreas Blattmann, Dominik Lorenz, Patrick Esser, and Bj{\"o}rn Ommer.
\newblock High-resolution image synthesis with latent diffusion models.
\newblock In \emph{Proceedings of the IEEE/CVF conference on computer vision and pattern recognition}, pages 10684--10695, 2022.

\bibitem[Roth et~al.(2015)Roth, Lu, Farag, Shin, Liu, Turkbey, and Summers]{roth2015deeporgan}
Holger~R Roth, Le Lu, Amal Farag, Hoo-Chang Shin, Jiamin Liu, Evrim~B Turkbey, and Ronald~M Summers.
\newblock Deeporgan: Multi-level deep convolutional networks for automated pancreas segmentation.
\newblock In \emph{International conference on medical image computing and computer-assisted intervention}, pages 556--564. Springer, 2015.

\bibitem[Saharia et~al.(2022)Saharia, Chan, Saxena, Li, Whang, Denton, Ghasemipour, Gontijo~Lopes, Karagol~Ayan, Salimans, et~al.]{saharia2022photorealistic}
Chitwan Saharia, William Chan, Saurabh Saxena, Lala Li, Jay Whang, Emily~L Denton, Kamyar Ghasemipour, Raphael Gontijo~Lopes, Burcu Karagol~Ayan, Tim Salimans, et~al.
\newblock Photorealistic text-to-image diffusion models with deep language understanding.
\newblock \emph{Advances in neural information processing systems}, 35:\penalty0 36479--36494, 2022.

\bibitem[Shin et~al.(2018)]{shin2018abnormal}
Younghak Shin et~al.
\newblock Abnormal colon polyp image synthesis using conditional adversarial networks for improved detection performance.
\newblock \emph{IEEE Access}, 6:\penalty0 56007--56017, 2018.

\bibitem[Wang et~al.(2022)Wang, Zhou, Zhang, Lei, Sun, Xu, and Xu]{wang2022anomaly}
Hualin Wang, Yuhong Zhou, Jiong Zhang, Jianqin Lei, Dongke Sun, Feng Xu, and Xiayu Xu.
\newblock Anomaly segmentation in retinal images with poisson-blending data augmentation.
\newblock \emph{Medical Image Analysis}, 81:\penalty0 102534, 2022.

\bibitem[Wei et~al.(2024{\natexlab{a}})Wei, Li, Fan, Ma, Qiu, Chen, and Lei]{wei2024sam2}
Jia Wei, Yun Li, Xiaomao Fan, Wenjun Ma, Meiyu Qiu, Hongyu Chen, and Wenbin Lei.
\newblock Sam-swin: Sam-driven dual-swin transformers with adaptive lesion enhancement for laryngo-pharyngeal tumor detection.
\newblock \emph{arXiv preprint arXiv:2410.21813}, 2024{\natexlab{a}}.

\bibitem[Wei et~al.(2024{\natexlab{b}})Wei, Li, Qiu, Chen, Fan, and Lei]{wei2024sam}
Jia Wei, Yun Li, Meiyu Qiu, Hongyu Chen, Xiaomao Fan, and Wenbin Lei.
\newblock Sam-fnet: Sam-guided fusion network for laryngo-pharyngeal tumor detection.
\newblock \emph{arXiv preprint arXiv:2408.05426}, 2024{\natexlab{b}}.

\bibitem[Wu et~al.(2024)Wu, Zhuang, Ni, and Chen]{wu2024freetumoradvancetumorsegmentation}
Linshan Wu, Jiaxin Zhuang, Xuefeng Ni, and Hao Chen.
\newblock Freetumor: Advance tumor segmentation via large-scale tumor synthesis, 2024.

\bibitem[Wyatt et~al.(2022)Wyatt, Leach, Schmon, and Willcocks]{wyatt2022anoddpm}
Julian Wyatt, Adam Leach, Sebastian~M Schmon, and Chris~G Willcocks.
\newblock Anoddpm: Anomaly detection with denoising diffusion probabilistic models using simplex noise.
\newblock In \emph{CVPR}, pages 650--656, 2022.

\bibitem[Xia et~al.(2022)Xia, Yu, Chu, Kawamoto, Park, Liu, Chen, Zhu, Li, Zhou, et~al.]{xia2022felix}
Yingda Xia, Qihang Yu, Linda Chu, Satomi Kawamoto, Seyoun Park, Fengze Liu, Jieneng Chen, Zhuotun Zhu, Bowen Li, Zongwei Zhou, et~al.
\newblock The felix project: Deep networks to detect pancreatic neoplasms.
\newblock \emph{medRxiv}, 2022.

\bibitem[Xiao et~al.(2022)Xiao, Bai, Yuille, and Zhou]{xiao2022delving}
Junfei Xiao, Yutong Bai, Alan Yuille, and Zongwei Zhou.
\newblock Delving into masked autoencoders for multi-label thorax disease classification.
\newblock \emph{IEEE Winter Conference on Applications of Computer Vision}, 2022.

\bibitem[Xiao et~al.(2025)Xiao, Zhou, Li, Lan, Mei, Yu, Zhao, Yuille, Zhou, and Xie]{xiao2025semantic}
Junfei Xiao, Ziqi Zhou, Wenxuan Li, Shiyi Lan, Jieru Mei, Zhiding Yu, Bingchen Zhao, Alan Yuille, Yuyin Zhou, and Cihang Xie.
\newblock A semantic space is worth 256 language descriptions: Make stronger segmentation models with descriptive properties.
\newblock In \emph{European Conference on Computer Vision}, pages 239--258. Springer, 2025.

\bibitem[Xu et~al.(2018)Xu, Zhang, Huang, Zhang, Gan, Huang, and He]{xu2018attngan}
Tao Xu, Pengchuan Zhang, Qiuyuan Huang, Han Zhang, Zhe Gan, Xiaolei Huang, and Xiaodong He.
\newblock Attngan: Fine-grained text to image generation with attentional generative adversarial networks.
\newblock In \emph{Proceedings of the IEEE conference on computer vision and pattern recognition}, pages 1316--1324, 2018.

\bibitem[Xu et~al.(2024)Xu, Sun, Peng, Jia, Morrison, Perer, Zandifar, Visweswaran, Eslami, and Batmanghelich]{xu2024medsyn}
Yanwu Xu, Li Sun, Wei Peng, Shuyue Jia, Katelyn Morrison, Adam Perer, Afrooz Zandifar, Shyam Visweswaran, Motahhare Eslami, and Kayhan Batmanghelich.
\newblock Medsyn: Text-guided anatomy-aware synthesis of high-fidelity 3d ct images.
\newblock \emph{IEEE Transactions on Medical Imaging}, 2024.

\bibitem[Yang et~al.(2024)Yang, Xu, Kang, Shi, and Zhao]{yang2024freemask}
Lihe Yang, Xiaogang Xu, Bingyi Kang, Yinghuan Shi, and Hengshuang Zhao.
\newblock Freemask: Synthetic images with dense annotations make stronger segmentation models.
\newblock \emph{Advances in Neural Information Processing Systems}, 36, 2024.

\bibitem[Yao et~al.(2021)Yao, Xiao, Liu, and Zhou]{yao2021label}
Qingsong Yao, Li Xiao, Peihang Liu, and S~Kevin Zhou.
\newblock Label-free segmentation of covid-19 lesions in lung ct.
\newblock \emph{IEEE Trans. Medical Imag.}, 40\penalty0 (10):\penalty0 2808--2819, 2021.

\bibitem[Yao et~al.(2024)Yao, Liu, Yin, Cheung, and Qin]{yao2024addressing}
Wenfang Yao, Chen Liu, Kejing Yin, William~K Cheung, and Jing Qin.
\newblock Addressing asynchronicity in clinical multimodal fusion via individualized chest x-ray generation.
\newblock \emph{arXiv preprint arXiv:2410.17918}, 2024.

\bibitem[Zhang et~al.(2019)Zhang, Xie, Xia, and Shen]{zhang2019lung}
Jun Zhang, Yuting Xie, Yong Xia, and Chunhua Shen.
\newblock Lung nodule classification with multi-scale convolutional neural networks.
\newblock In \emph{International Conference on Medical Image Computing and Computer-Assisted Intervention}, pages 588--595. Springer, 2019.

\bibitem[Zhang et~al.(2023)Zhang, Guo, and Jin]{zhang2023radiomics}
Wenchao Zhang, Yu Guo, and Qiyu Jin.
\newblock Radiomics and its feature selection: A review.
\newblock \emph{Symmetry}, 15\penalty0 (10):\penalty0 1834, 2023.

\end{thebibliography}
}

\clearpage
\appendix
\setcounter{page}{1}
\onecolumn
\renewcommand \thepart{}
\renewcommand \partname{}
\part{Appendix} 
\setcounter{secnumdepth}{4}
\setcounter{tocdepth}{4}
\parttoc 

\clearpage
\section{Text-Driven Visual Examples}\label{Visual_Examples}
\subsection{Text-Driven Pancreas Classification Visual Examples}\label{sec:supp_Pancreas_Classification}
There are three common types of tumors in the pancreas: Pancreatic Ductal Adenocarcinoma (PDAC), Pancreatic Neuroendocrine Tumors (PNET), and cystic tumors.

\smallskip\noindent\textbf{PDAC (Pancreatic Ductal Adenocarcinoma)}: This is the most common and aggressive pancreatic tumor, arising from the ductal cells. On imaging, PDAC typically appears as a \textit{poorly defined, hypoattenuating (dark) mass with minimal contrast enhancement}. Its subtle imaging features and late clinical presentation contribute to its poor prognosis.

\smallskip\noindent\textbf{PNET (Pancreatic Neuroendocrine Tumors)}: PNET are less common and originate from the endocrine cells of the pancreas. A key characteristic of PNET is their \textit{bright appearance on contrast-enhanced imaging, particularly during the arterial phase, due to their hypervascularity}. These well-defined, hyperenhancing lesions stand out against the relatively lower-density pancreatic tissue. This brightness is a critical diagnostic feature and highlights their vascular nature. Functional PNET may cause hormone-related syndromes, while non-functional ones are often detected incidentally.

\smallskip\noindent\textbf{Cystic Tumors}: Pancreatic cystic neoplasms, such as serous cystadenomas, mucinous cystic neoplasms (MCNs), and intraductal papillary mucinous neoplasms (IPMNs), are \textit{fluid-filled lesions that may have internal septations or solid components.} While some are benign, others carry malignant potential and require careful evaluation.

\begin{figure}[ht]
  \centering
\includegraphics[width=0.63\linewidth]{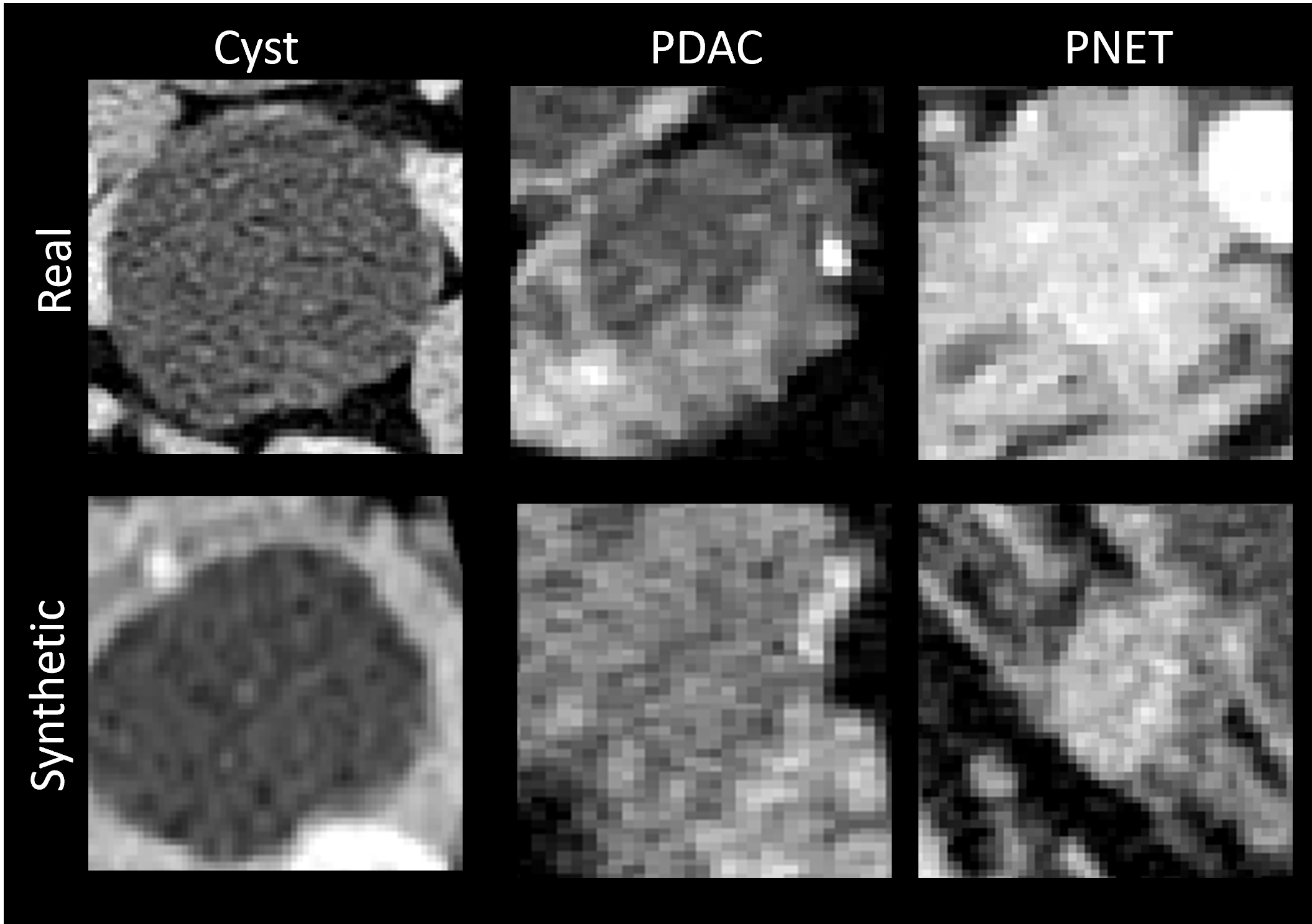}
    \caption{Comparison of real and synthetic pancreatic tumors across three types: Cyst, PDAC, and PNET. The top row displays real medical images, highlighting the distinct characteristics of each tumor type—cysts with smooth, fluid-filled appearances, hypoattenuating PDAC masses, and hypervascular PNET lesions with bright enhancement. The bottom row shows their corresponding synthetic counterparts, demonstrating the ability of the model to replicate texture, shape, and contrast features unique to each tumor type.}
    \label{fig:sup_classification}
\end{figure}

\begin{table}[ht]
\centering
\scriptsize
\begin{tabular}{p{0.06\linewidth}p{0.1\linewidth}p{0.08\linewidth}p{0.08\linewidth}p{0.08\linewidth}|p{0.08\linewidth}p{0.08\linewidth}p{0.08\linewidth}}
\toprule
Level & Method & \multicolumn{3}{c|}{Malignant Tumor} & \multicolumn{3}{c}{Benign Cyst} \\
\cmidrule{3-8}
& & Sen & Spec & PPV & Sen & Spe & PPV \\
\midrule
\multirow{2}{*}{Patient} & RealTumor & 80.1 \scalebox{.6}{(347/433)} & 70.1 \scalebox{.6}{(373/532)} & 68.6 \scalebox{.6}{(347/506)} & 77.8 \scalebox{.6}{(189/243)} & 61.9 \scalebox{.6}{(447/722)} & 40.7 \scalebox{.6}{(189/464)} \\
& \method & 81.5 \scalebox{.6}{(353/433)} & 73.3 \scalebox{.6}{(390/532)} & 71.3 \scalebox{.6}{(353/495)} & 84.0 \scalebox{.6}{(204/243)} & 77.1 \scalebox{.6}{(557/722)} & 55.3 \scalebox{.6}{(204/369)} \\
\midrule
& & Sen & DSC & NSD & Sen & DSC & NSD \\
\cmidrule{3-8}
\multirow{2}{*}{Tumor} & RealTumor & 61.9 \scalebox{.6}{(304/491)} & 28.1 & 24.7 & 50.7 \scalebox{.6}{(245/483)} & 39.6 & 43.5 \\
& \method & 70.1 \scalebox{.6}{(344/491)} & 45.5 & 40.7 & 57.8 \scalebox{.6}{(279/483)}  & 42.1 & 49.2 \\
\bottomrule
\end{tabular}
\caption{\textbf{Patient- and Tumor-Level of pancreatic tumor classification.} Patient-level metrics evaluate the model's ability to detect tumors based on CT scans, using sensitivity, specificity, and PPV. Tumor-level metrics assess localization and segmentation through sensitivity, Dice similarity coefficient (DSC), and normalized surface distance (NSD), considering both malignant and benign tumors.}
\label{tab:sup_tumor_classification}
\end{table}

\clearpage
\subsection{Early Detection cases}\label{sec:supp_Early_Detection}
\method\ leverages descriptive text to contextualize each tumor's visual and structural features, enabling the model to refine its understanding of nuanced patterns. This approach utilizes a high-frequency diagnostic vocabulary to generate text descriptions that align with each tumor's visual characteristics. By incorporating this text-driven guidance into the detection process, \method\ achieves improved sensitivity and specificity, particularly in identifying difficult cases.
\begin{figure}[ht]
  \centering
\includegraphics[width=\linewidth]{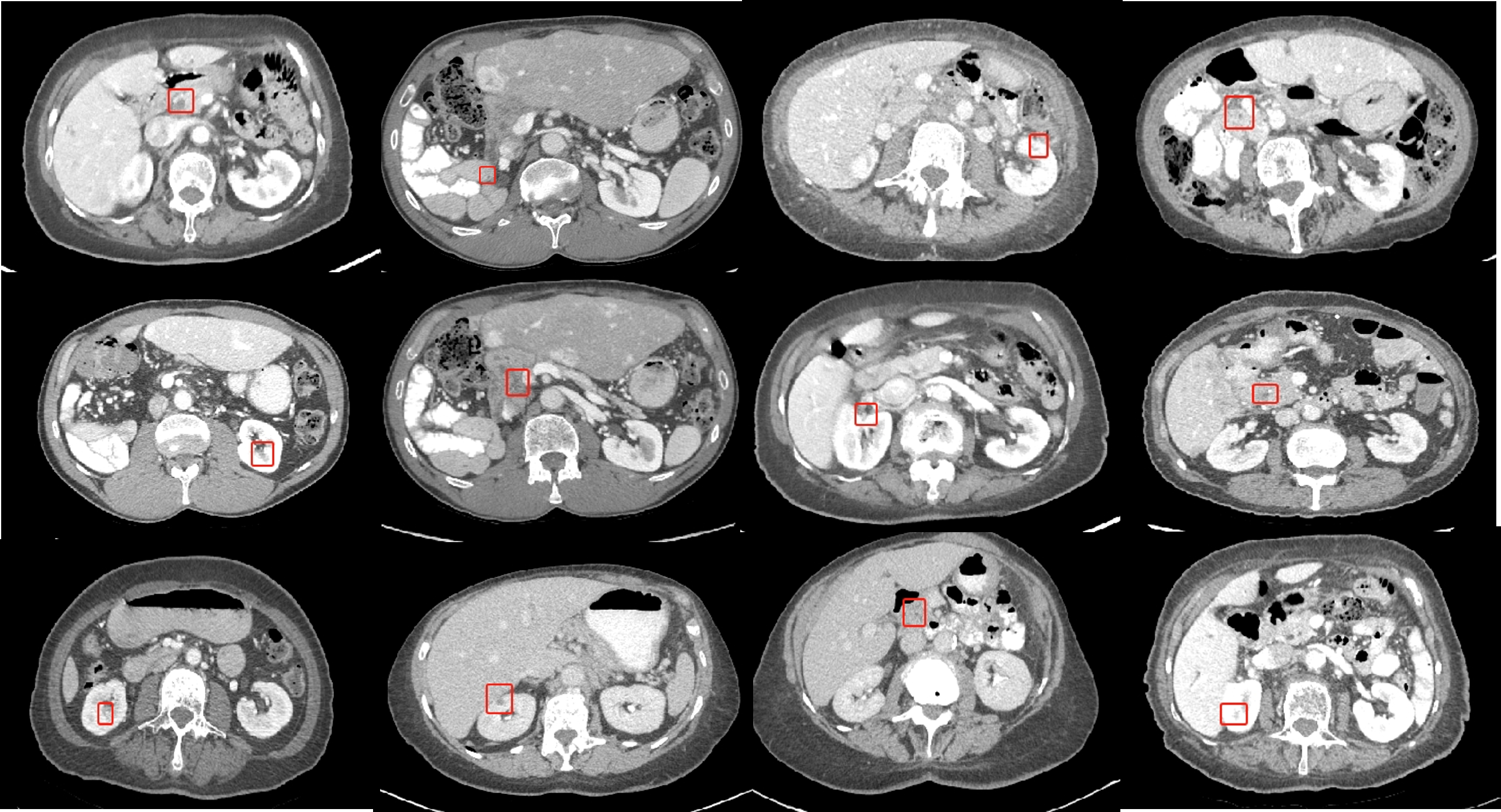}
    \caption{Text-driven approaches enhance early tumor detection by identifying subtle cases overlooked by DiffTumor. The results demonstrate \method's ability to leverage descriptive text to improve detection accuracy, particularly in challenging scenarios.}
    \label{fig:Early_Detection}
\end{figure}
\clearpage
\subsection{Text-Driven Targeted Data Visual Examples}\label{sec:supp_Targeted_Data_Visual_Examples}
\begin{figure}[ht]
  \centering
\includegraphics[width=0.57\linewidth]{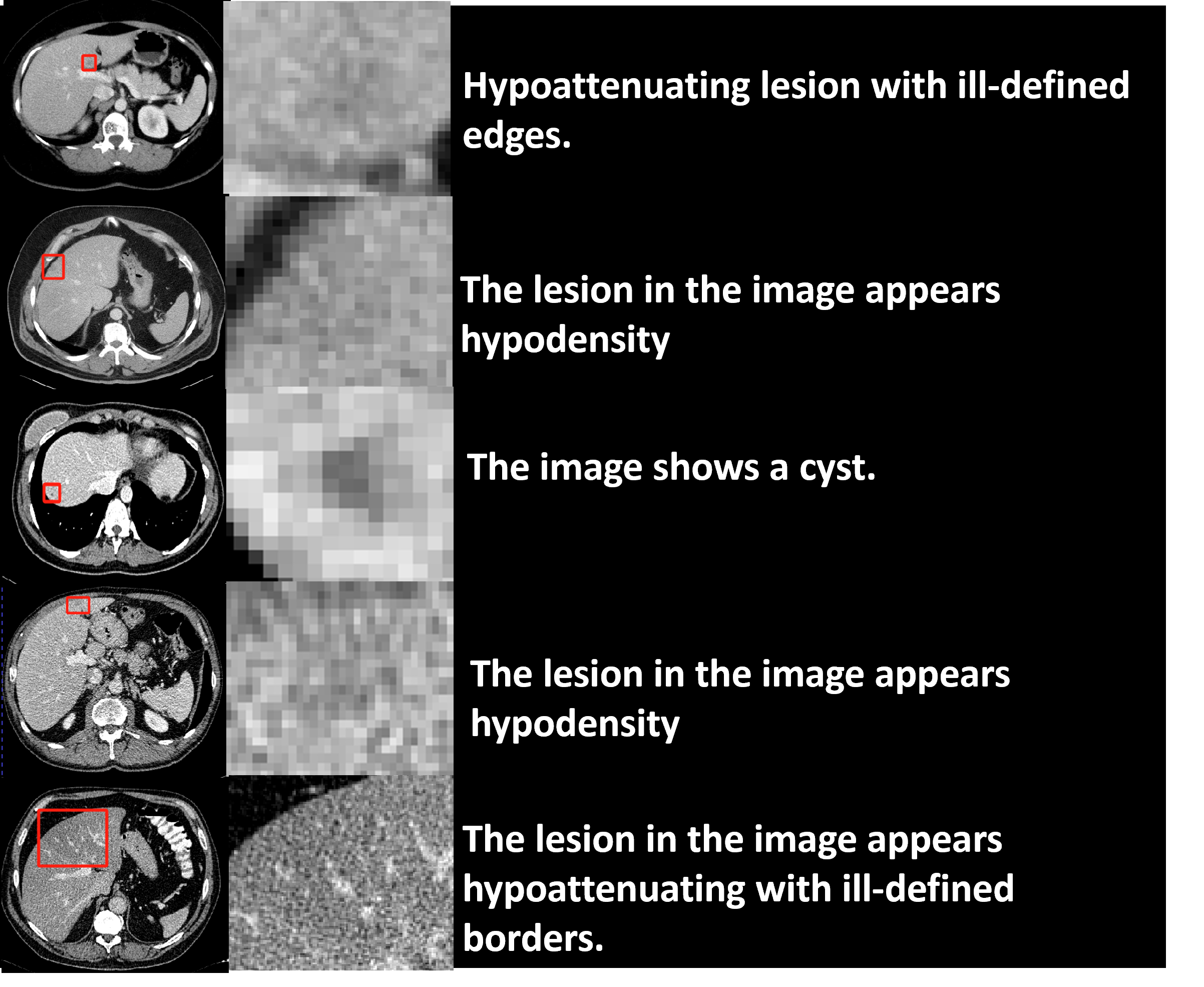}
    \caption{\textbf{FN cases with descriptive words}: False-negative (FN) tumor cases undetected by the baseline DiffTumor model, paired with descriptive words generated by GPT-4o. The descriptions, based on high-frequency diagnostic terminology, highlight key tumor characteristics such as texture, margin irregularities, and enhancement patterns. These descriptive words provide essential context for downstream tumor synthesis and model augmentation.}
    \label{fig:Targeted_Data_Visual_Examples}
\end{figure}
\begin{figure}[ht]
  \centering
\includegraphics[width=0.57\linewidth]{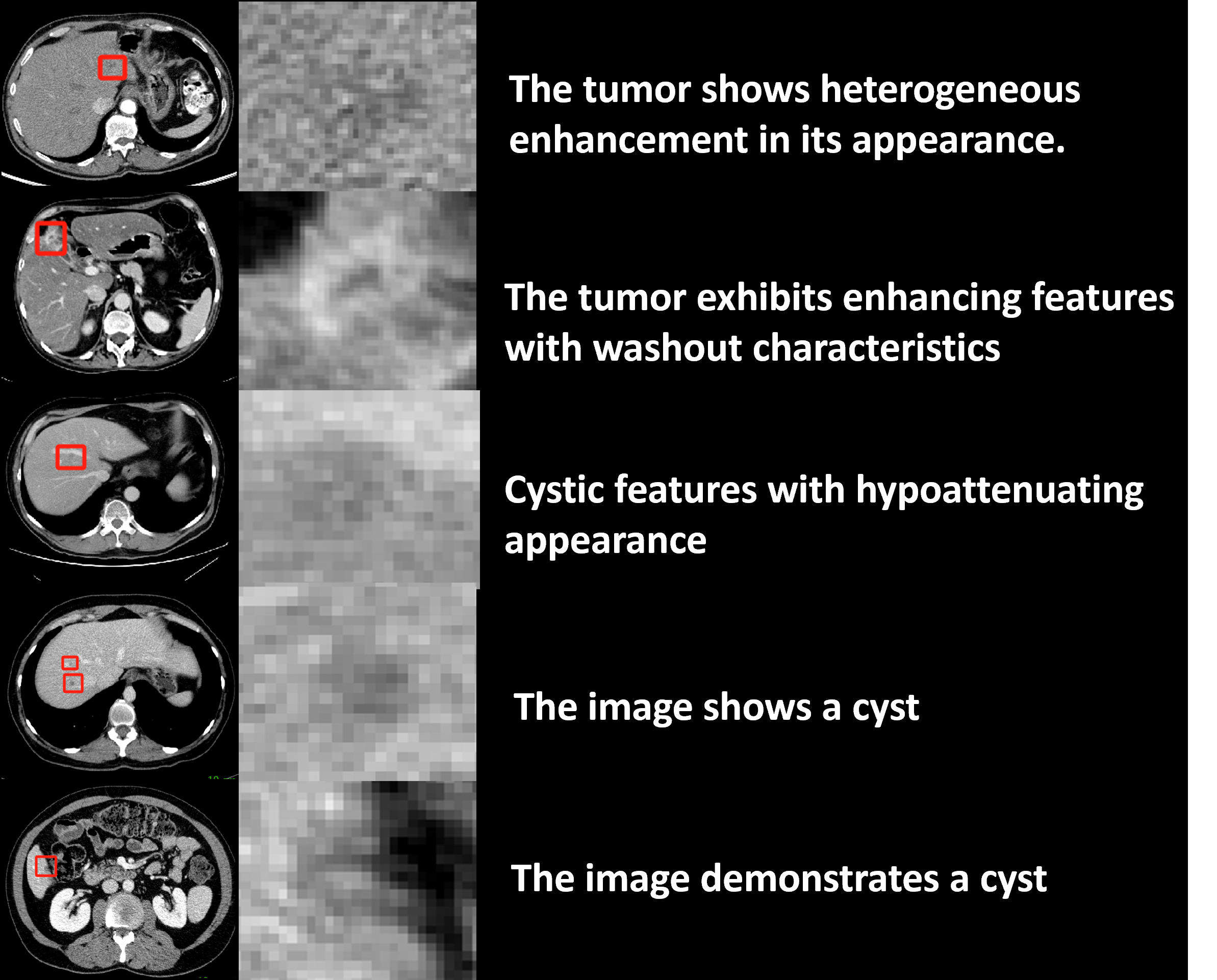}
    \caption{\textbf{Text-Driven Targeted Data}: Tumor instances synthesized using GPT-4o-generated descriptive words and corresponding magnified tumor masks. The synthetic tumors replicate challenging features, including subtle textures and indistinct boundaries, enhancing the training dataset. This integration of descriptive text, zoomed-in masks, and healthy CT scans contributes to improved model sensitivity and detection accuracy, addressing limitations in identifying complex tumor presentations.}
    \label{fig:classification}
\end{figure}

\clearpage
\section{Ablation Study}\label{sec:supp_Ablation}
\subsection{Overall}\label{sec:supp_Ablation_overall}
In this experiment, we utilized healthy CT data to synthesize tumors. \method\ (with all components), and its variants excluding Text-Driven Contrastive Learning (Contrastive Loss), Text Extraction and Generation (Text E-G), and Targeted Data Augmentation (T-D-A), as well as the baseline model DiffTumor and the real tumor data (RealTumor). The Healthy dataset was paired with real tumors in about 1:1 ratio to probabilistically generate tumors of varying sizes. Due to GPU resource constraints, we present results using only fold 0 and fold 1.

\begin{table*}[ht]
\centering
\scriptsize
\begin{tabular}{p{0.15\linewidth}|p{0.1\linewidth} p{0.1\linewidth} p{0.1\linewidth}|c|c|c|c|c}
\toprule
\textbf{Method} &&&& \multicolumn{3}{c}{\textbf{Tumor Size (d, mm)}}  & \textbf{DSC (\%)} & \textbf{NSD (\%)} \\
& Text E-G & Contrastive Loss & T-D-A & $d<20$ & $20 \leq d < 50$ & $d \geq 50$ & & \\
\midrule
\multicolumn{9}{c}{\textbf{Liver}} \\
\midrule
RealTumor & - & - & -
& 64.5\scalebox{.6}{ (20/31)} & 69.7\scalebox{.6}{ (53/76)} & 66.7\scalebox{.6}{ (38/57)}  & 59.1\scalebox{.6}{$\pm30.4$} & 60.1\scalebox{.6}{$\pm30.0$} \\
SynTumor~\cite{hu2023label}& - & - & -
& 71.0\scalebox{.6}{ (22/31)} & 69.7\scalebox{.6}{ (53/76)} & 73.7\scalebox{.6}{ (42/57)}  & 62.3\scalebox{.6}{$\pm12.7$} & 87.7\scalebox{.6}{$\pm21.4$} \\
Pixel2Cancer~\cite{lai2024pixel}& - & - & -
& - & - & -  & 57.2\scalebox{.6}{$\pm21.3$} & 63.1\scalebox{.6}{$\pm15.6$} \\
DiffTumor~\cite{chen2024towards}& - & - & -
& 77.4\scalebox{.6}{ (24/31)} & 75.0\scalebox{.6}{ (57/76)} & 73.7\scalebox{.6}{ (42/57)}  & 64.2\scalebox{.6}{$\pm33.3$} & 66.1\scalebox{.6}{$\pm32.8$} \\
\method\ &\ding{55}&\ding{55}&\ding{55}
& 75.4\scalebox{.6}{ (23/31)} & 72.4\scalebox{.6}{ (55/76)} & 74.6\scalebox{.6}{ (43/57)}  & 65.5\scalebox{.6}{$\pm25.0$} & 61.3\scalebox{.6}{$\pm28.6$} \\
\method\ &\ding{51}&\ding{55}&\ding{55}
& 77.4\scalebox{.6}{ (24/31)} & 75.0\scalebox{.6}{ (57/76)} & 77.2\scalebox{.6}{ (44/57)}  & 68.4\scalebox{.6}{$\pm30.4$} & 69.2\scalebox{.6}{$\pm31.0$} \\
\method\ &\ding{51}&\ding{51}&\ding{55}
& 80.6\scalebox{.6}{ (25/31)} & 77.6\scalebox{.6}{ (59/76)} & 80.7\scalebox{.6}{ (46/57)}  & 69.7\scalebox{.6}{$\pm27.2$} & 70.8\scalebox{.6}{$\pm26.0$} \\
\method\ &\ding{51}&\ding{51}&\ding{51}
& 83.9\scalebox{.6}{ (26/31)} & 77.6\scalebox{.6}{ (59/76)} & 87.7\scalebox{.6}{ (50/57)}  & 71.6\scalebox{.6}{$\pm27.2$} & 72.4\scalebox{.6}{$\pm30.3$} \\
\midrule
\multicolumn{9}{c}{\textbf{Pancreas}} \\
\midrule
RealTumor& - & - & -
& 58.3\scalebox{.6}{ (14/24)} & 67.7\scalebox{.6}{ (21/31)} & 57.1\scalebox{.6}{ (4/7)} & 53.3\scalebox{.6}{$\pm28.7$} & 40.1\scalebox{.6}{$\pm28.8$} \\
SynTumor~\cite{hu2023label}& - & - & -
& 62.5\scalebox{.6}{ (15/24)} & 64.5\scalebox{.6}{ (20/31)} & 57.1\scalebox{.6}{ (4/7)} & 54.0\scalebox{.6}{$\pm31.4$} & 47.2\scalebox{.6}{$\pm23.0$} \\
Pixel2Cancer~\cite{lai2024pixel}& - & - & -
& - & - & - & 57.9\scalebox{.6}{$\pm13.7$} & 54.3\scalebox{.6}{$\pm19.2$} \\
DiffTumor~\cite{chen2024towards}& - & - & -
& 66.7\scalebox{.6}{ (16/24)} & 67.7\scalebox{.6}{ (21/31)} & 57.1\scalebox{.6}{ (4/7)}  & 58.9\scalebox{.6}{$\pm42.8$} & 52.8\scalebox{.6}{$\pm26.2$} \\
\method\ &\ding{55}&\ding{55}&\ding{55}
& 66.7\scalebox{.6}{ (16/24)} & 64.5\scalebox{.6}{ (20/31)} & 57.1\scalebox{.6}{ (4/7)}  & 55.8\scalebox{.6}{$\pm32.6$} & 51.1\scalebox{.6}{$\pm35.6$} \\
\method\ &\ding{51}&\ding{55}&\ding{55}
& 70.8\scalebox{.6}{ (17/24)} & 61.3\scalebox{.6}{ (19/31)} & 57.1\scalebox{.6}{ (4/7)}  & 59.7\scalebox{.6}{$\pm36.1$} & 60.6\scalebox{.6}{$\pm38.3$} \\
\method\ &\ding{51}&\ding{51}&\ding{55}
& 64.0\scalebox{.6}{ (16/24)} & 70.0\scalebox{.6}{ (21/31)} & 57.1\scalebox{.6}{ (4/7)}  & 60.2\scalebox{.6}{$\pm27.3$} & 71.0\scalebox{.6}{$\pm31.5$} \\
\method\ &\ding{51}&\ding{51}&\ding{51}
& 87.5\scalebox{.6}{ (21/24)} & 87.1\scalebox{.6}{ (27/31)} & 85.7\scalebox{.6}{ (6/7)}  & 67.3\scalebox{.6}{$\pm24.8$} & 65.5\scalebox{.6}{$\pm27.1$} \\
\midrule
\multicolumn{9}{c}{\textbf{Kidney}} \\
\midrule
RealTumor & - & - & -
& 71.4\scalebox{.6}{ (5/7)} & 66.7\scalebox{.6}{ (4/6)} & 69.0\scalebox{.6}{ (29/42)}  & 78.0\scalebox{.6}{$\pm14.9$} & 65.8\scalebox{.6}{$\pm17.7$} \\
SynTumor~\cite{hu2023label}& - & - & -
& 71.4\scalebox{.6}{ (5/7)} & 66.7\scalebox{.6}{ (4/6)} & 69.0\scalebox{.6}{ (29/42)}  & 78.1\scalebox{.6}{$\pm23.0$} & 66.0\scalebox{.6}{$\pm21.2$} \\
Pixel2Cancer~\cite{lai2024pixel}& - & - & -
& - & - & -  & 71.5\scalebox{.6}{$\pm21.4$} & 64.3\scalebox{.6}{$\pm16.9$} \\
DiffTumor~\cite{chen2024towards}& - & - & -
& 71.4\scalebox{.6}{ (5/7)} & 83.3\scalebox{.6}{ (5/6)} & 69.0\scalebox{.6}{ (29/42)}  & 78.9\scalebox{.6}{$\pm19.7$} & 69.2\scalebox{.6}{$\pm18.5$} \\
\method\ &\ding{55}&\ding{55}&\ding{55}
& 57.1\scalebox{.6}{ (4/7)} & 83.3\scalebox{.6}{ (5/6)} & 69.0\scalebox{.6}{ (29/42)} & 79.2\scalebox{.6}{$\pm22.3$} & 71.4\scalebox{.6}{$\pm21.4$} \\
\method\ &\ding{51}&\ding{55}&\ding{55}
& 71.4\scalebox{.6}{ (5/7)} & 83.3\scalebox{.6}{ (5/6)} & 76.2\scalebox{.6}{ (32/42)}  & 80.6\scalebox{.6}{$\pm21.8$} & 76.8\scalebox{.6}{$\pm19.3$} \\
\method\ &\ding{51}&\ding{51}&\ding{55}
& 71.4\scalebox{.6}{ (5/7)} & 83.3\scalebox{.6}{ (5/6)} & 73.8\scalebox{.6}{ (31/42)}  & 79.7\scalebox{.6}{$\pm20.2$} & 75.2\scalebox{.6}{$\pm21.5$} \\
\method\ &\ding{51}&\ding{51}&\ding{51}
& 71.4\scalebox{.6}{ (5/7)} & 83.3\scalebox{.6}{ (5/6)} & 76.2\scalebox{.6}{ (32/42)} & 85.2\scalebox{.6}{$\pm9.7$} & 78.4\scalebox{.6}{$\pm13.9$} \\
\bottomrule
\end{tabular}
\caption{\textbf{Ablation Study/fold 0}: Comparison of sensitivity (Sen\%), specificity (Spe\%), Dice Similarity Coefficient (DSC\%), and Normalized Surface Distance (NSD\%) for liver, pancreas, and kidney tumors using synthetic data for training with U-Net.}
\label{sup_Ablation_Study1}
\end{table*}

\begin{table*}[ht]
\centering
\scriptsize
\begin{tabular}{p{0.15\linewidth}|p{0.1\linewidth} p{0.1\linewidth} p{0.1\linewidth}|c|c|c|c|c}
\toprule
\textbf{Method} &&&& \multicolumn{3}{c}{\textbf{Tumor Size (d, mm)}}  & \textbf{DSC (\%)} & \textbf{NSD (\%)} \\
& Text E-G & Contrastive Loss & T-D-A & $d<20$ & $20 \leq d < 50$ & $d \geq 50$ & & \\
\midrule
\multicolumn{9}{c}{\textbf{Liver}} \\
\midrule
RealTumor& - & - & -
& 71.9\scalebox{.6}{ (23/32)} & 68.0\scalebox{.6}{ (51/75)} & 68.4\scalebox{.6}{ (39/57)}  & 60.2\scalebox{.6}{$\pm21.3$} & 63.5\scalebox{.6}{$\pm27.8$} \\
SynTumor~\cite{hu2023label}& - & - & -
& 84.4\scalebox{.6}{ (27/32)} & 81.3\scalebox{.6}{ (61/75)} & 78.9\scalebox{.6}{ (45/57)}  & 68.2\scalebox{.6}{$\pm14.0$} & 78.1\scalebox{.6}{$\pm16.7$} \\
Pixel2Cancer~\cite{lai2024pixel}& - & - & -
& - & - & -  & 60.3\scalebox{.6}{$\pm21.5$} & 62.0\scalebox{.6}{$\pm19.4$} \\
DiffTumor~\cite{chen2024towards}& - & - & -
& 81.3\scalebox{.6}{ (26/32)} & 77.3\scalebox{.6}{ (58/75)} & 82.5\scalebox{.6}{ (47/57)}  & 70.3\scalebox{.6}{$\pm23.1$} & 69.9\scalebox{.6}{$\pm36.1$} \\
\method\ &\ding{55}&\ding{55}&\ding{55}
& 75.0\scalebox{.6}{ (24/32)} & 76.0\scalebox{.6}{ (57/75)} & 77.2\scalebox{.6}{ (44/57)}  & 67.5\scalebox{.6}{$\pm18.9$} & 66.0\scalebox{.6}{$\pm21.7$} \\
\method\ &\ding{51}&\ding{55}&\ding{55}
& 78.1\scalebox{.6}{ (25/32)} & 80.0\scalebox{.6}{ (60/75)} & 78.9\scalebox{.6}{ (45/57)}  & 69.5\scalebox{.6}{$\pm21.4$} & 71.1\scalebox{.6}{$\pm29.9$} \\
\method\ &\ding{51}&\ding{51}&\ding{55}
& 81.3\scalebox{.6}{ (26/32)} & 80.0\scalebox{.6}{ (60/75)} & 87.7\scalebox{.6}{ (50/57)}  & 70.4\scalebox{.6}{$\pm26.6$} & 73.7\scalebox{.6}{$\pm28.7$} \\
\method\ &\ding{51}&\ding{51}&\ding{51}
& 90.6\scalebox{.6}{ (29/32)} & 92.0\scalebox{.6}{ (69/75)} & 94.7\scalebox{.6}{ (54/57)}  & 75.4\scalebox{.6}{$\pm19.3$} & 76.6\scalebox{.6}{$\pm22.9$} \\
\midrule
\multicolumn{9}{c}{\textbf{Pancreas}} \\
\midrule
RealTumor& - & - & -
& 68.0\scalebox{.6}{ (17/25)} & 76.7\scalebox{.6}{ (23/30)} & 33.3\scalebox{.6}{ (1/3)}  & 55.2\scalebox{.6}{$\pm18.4$} & 47.3\scalebox{.6}{$\pm24.1$} \\
SynTumor~\cite{hu2023label}& - & - & -
& 80.0\scalebox{.6}{ (20/25)} & 76.7\scalebox{.6}{ (23/30)} & 33.3\scalebox{.6}{ (1/3)}  & 56.3\scalebox{.6}{$\pm22.4$} & 49.1\scalebox{.6}{$\pm19.6$} \\
Pixel2Cancer~\cite{lai2024pixel}& - & - & -
& - & - & -  & 60.7\scalebox{.6}{$\pm26.6$} & 58.2\scalebox{.6}{$\pm13.2$} \\
DiffTumor~\cite{chen2024towards}& - & - & -
& 92.0\scalebox{.6}{ (23/25)} & 80.0\scalebox{.6}{ (24/30)} & 33.3\scalebox{.6}{ (1/3)}  & 59.0\scalebox{.6}{$\pm32.7$} & 60.6\scalebox{.6}{$\pm17.3$} \\
\method\ &\ding{55}&\ding{55}&\ding{55}
& 92.0\scalebox{.6}{ (23/25)} & 76.7\scalebox{.6}{ (23/30)} & 0.0\scalebox{.6}{ (0/3)}  & 57.2\scalebox{.6}{$\pm30.1$} & 50.8\scalebox{.6}{$\pm30.4$} \\
\method\ &\ding{51}&\ding{55}&\ding{55}
& 92.0\scalebox{.6}{ (23/25)} & 80.0\scalebox{.6}{ (24/30)} & 66.7\scalebox{.6}{ (2/3)}  & 62.1\scalebox{.6}{$\pm25.1$} & 67.3\scalebox{.6}{$\pm27.9$} \\
\method\ &\ding{51}&\ding{51}&\ding{55}
& 88.0\scalebox{.6}{ (22/25)} & 80.0\scalebox{.6}{ (24/30)} & 66.7\scalebox{.6}{ (2/3)}  & 64.3\scalebox{.6}{$\pm22.5$} & 69.8\scalebox{.6}{$\pm29.8$} \\
\method\ &\ding{51}&\ding{51}&\ding{51}
& 100.0\scalebox{.6}{ (25/25)} & 90.0\scalebox{.6}{ (27/30)} & 100.0\scalebox{.6}{ (3/3)}  & 69.6\scalebox{.6}{$\pm19.3$} & 73.2\scalebox{.6}{$\pm20.1$} \\
\midrule
\multicolumn{9}{c}{\textbf{Kidney}} \\
\midrule
RealTumor& - & - & -
& 50.0\scalebox{.6}{ (5/10)} & 60.0\scalebox{.6}{ (3/5)} & 65.9\scalebox{.6}{ (29/44)}  & 79.2\scalebox{.6}{$\pm14.2$} & 65.1\scalebox{.6}{$\pm11.3$} \\
SynTumor~\cite{hu2023label}& - & - & -
& 70.0\scalebox{.6}{ (7/10)} & 100.0\scalebox{.6}{ (5/5)} & 84.1\scalebox{.6}{ (37/44)}  & 80.3\scalebox{.6}{$\pm12.8$} & 72.9\scalebox{.6}{$\pm18.5$} \\
Pixel2Cancer~\cite{lai2024pixel}& - & - & -
& - & - & -  & 61.6\scalebox{.6}{$\pm22.8$} & 69.8\scalebox{.6}{$\pm14.0$} \\
DiffTumor~\cite{chen2024towards}& - & - & -
& 70.0\scalebox{.6}{ (7/10)} & 100.0\scalebox{.6}{ (5/5)} & 81.8\scalebox{.6}{ (36/44)}  & 80.4\scalebox{.6}{$\pm19.7$} & 79.7\scalebox{.6}{$\pm9.2$} \\
\method\ &\ding{55}&\ding{55}&\ding{55}
& 70.0\scalebox{.6}{ (7/10)} & 100.0\scalebox{.6}{ (5/5)} & 86.4\scalebox{.6}{ (38/44)} & 81.3\scalebox{.6}{$\pm17.7$} & 78.4\scalebox{.6}{$\pm16.4$} \\
\method\ &\ding{51}&\ding{55}&\ding{55}
& 70.0\scalebox{.6}{ (7/10)} & 100.0\scalebox{.6}{ (5/5)} & 84.1\scalebox{.6}{ (37/44)}  & 80.9\scalebox{.6}{$\pm24.0$} & 79.3\scalebox{.6}{$\pm21.4$} \\
\method\ &\ding{51}&\ding{51}&\ding{55}
& 70.0\scalebox{.6}{ (7/10)} & 100.0\scalebox{.6}{ (5/5)} & 86.4\scalebox{.6}{ (38/44)}  & 82.0\scalebox{.6}{$\pm18.2$} & 80.2\scalebox{.6}{$\pm14.9$} \\
\method\ &\ding{51}&\ding{51}&\ding{51}
& 90.0\scalebox{.6}{ (9/10)} & 100.0\scalebox{.6}{ (5/5)} & 95.5\scalebox{.6}{ (42/44)} & 86.7\scalebox{.6}{$\pm12.3$} & 82.9\scalebox{.6}{$\pm19.4$} \\
\bottomrule
\end{tabular}
\caption{\textbf{Ablation Study/fold 1}: Comparison of sensitivity (Sen\%), specificity (Spe\%), Dice Similarity Coefficient (DSC\%), and Normalized Surface Distance (NSD\%) for liver, pancreas, and kidney tumors using synthetic data for training with U-Net.}
\label{sup_Ablation_Study2}
\end{table*}
\clearpage
\subsection{Text Extraction and Generation}
\label{sec:supp_text_extraction}
\begin{figure*}[ht]
    \centering
    \includegraphics[width=\linewidth]{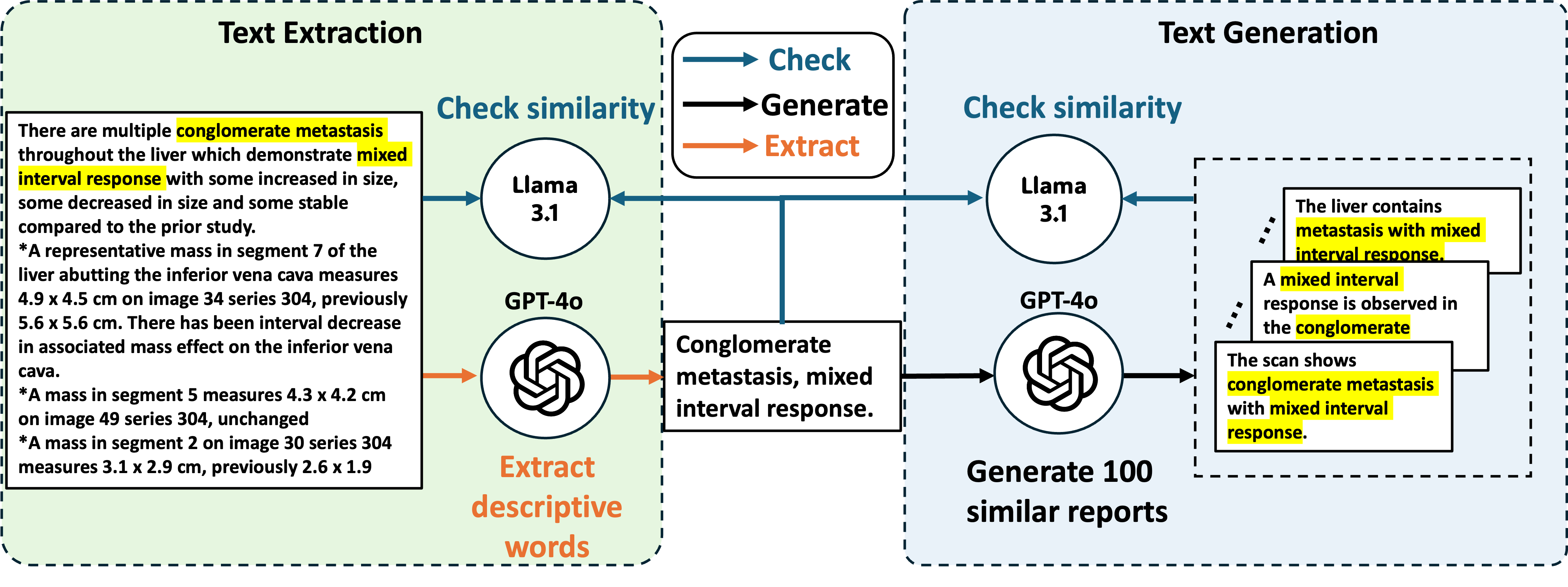}
    \caption{\textbf{Text Extraction and Generation}: We illustrate a two-step workflow for radiology reports, divided into Text Extraction and Text Generation. In the \textbf{Text Extraction}, complex radiology reports are processed with GPT-4o to extract descriptive words, capturing essential details such as `\textit{mixed interval response.}' These extracted descriptors are compared to the original report to ensure descriptive alignment. In the \textbf{Text Generation}, the extracted descriptive words are used to generate 100 similar reports, each maintaining core descriptive details while varying in structure. These generated reports are further assessed for similarity with the extracted descriptors, ensuring consistent descriptive content throughout the augmented data.}
    \label{fig:dataaugmentation}
\end{figure*}

In current radiology reports, controlling tumor synthesis through textual descriptions faces numerous challenges. (1) the reports often contain \textbf{substantial irrelevant or noisy information}, and tumor characteristics (such as shape, size, and location) are frequently presented in a fragmented manner. These characteristics are often more effectively obtained and managed through tumor masks. For example, when describing that `\textit{there are multiple conglomerate metastases throughout the liver demonstrating mixed interval response with some increased in size, some decreased, and others stable compared to the prior study,}' terms like `\textit{conglomerate metastases}' and `\textit{mixed interval response}' serve as key descriptive features, while other details may be considered noise; (2) descriptions often exhibit \textbf{discontinuity and inconsistency}~\cite{xiao2025semantic}. A particular tumor characteristic may be scattered across multiple sentences or paragraphs, making it insufficient to rely on a single descriptive term to fully represent it. Instead, one should enhance the representation of the tumor's characteristics by incorporating multiple similar descriptions expressed in a variety of sentence structures. By employing multiple, similarly meaningful but differently phrased descriptions, the consistency, reliability, and representativeness of the textual data can be improved, thus facilitating more robust automated analysis and feature extraction. These issues undermine the efficacy of radiology reports as reliable conditions in diffusion models for tumor generation. To address these challenges and achieve more controlled tumor synthesis via text descriptions, we implemented a two-stage data preprocessing approach, including data cleaning and data augmentation, as shown in \figureautorefname~\ref{fig:dataaugmentation}.

\clearpage
\subsection{Targeted Data Augmentation}
\label{sec:supp_targeted}
\begin{figure}[ht]
  \centering
\includegraphics[width=0.6\linewidth]{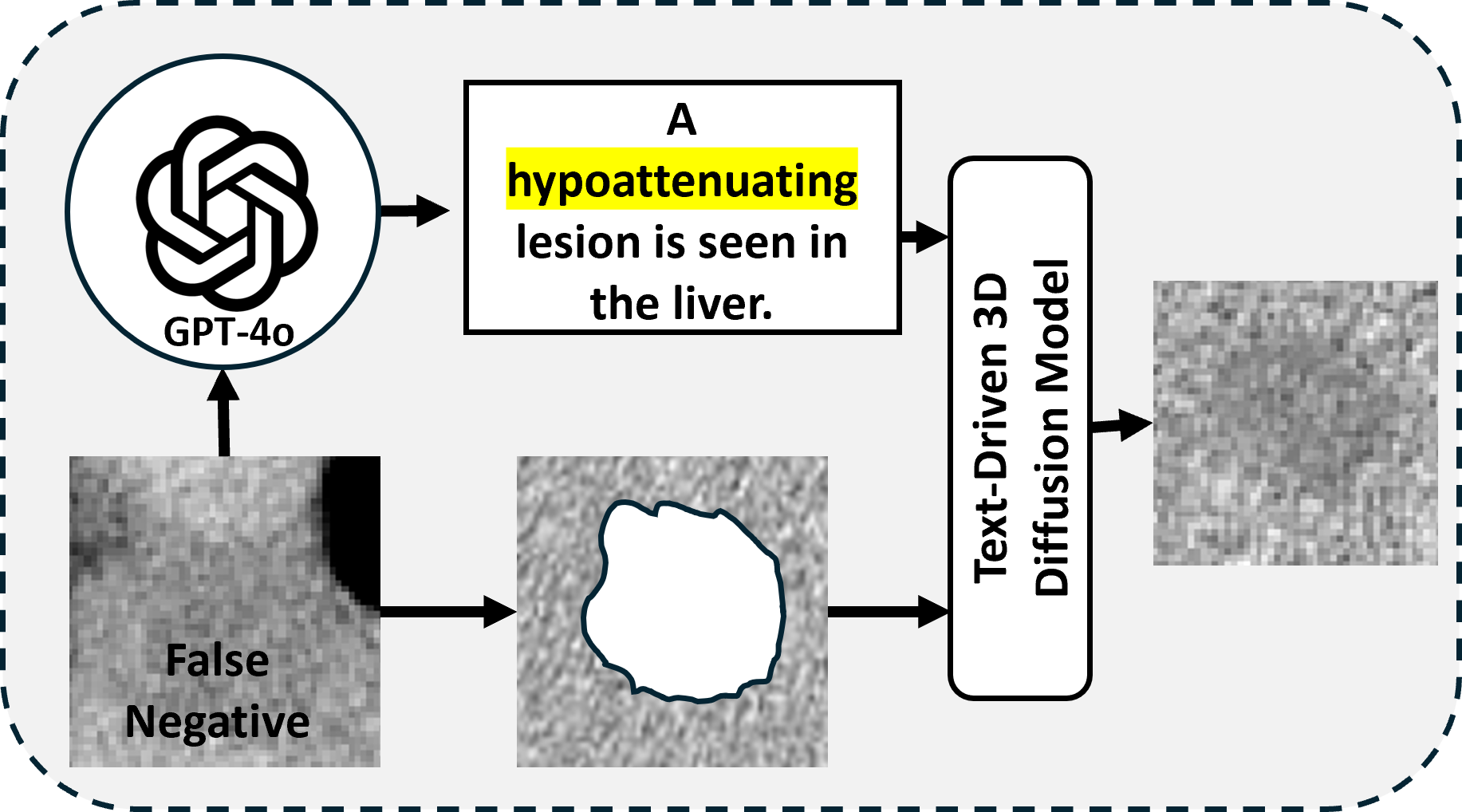}
    \caption{\textbf{Targeted Data Augmentation.} This workflow aims to improve prior art's detection of previously False Positive. The input consists of a missed case's tumor mask, a healthy CT scan, and a descriptive text generated by GPT-4o (e.g., `\textit{a hypoattenuating lesion is seen in the liver}'). This structured input is processed by the diffusion model to output synthetic tumor images, enriching the dataset and enhancing detection accuracy.}
    \label{fig:Adversarial}
\end{figure}
To study the contribution of Targeted Data Augmentation, as shown in \figureautorefname~\ref{fig:Adversarial}, particularly in cases where prior approaches have shown limitations, we we introduce Targeted Data Augmentation by leveraging False Positive tumor. We first collected a related paired dataset for these challenging examples, including the CT scans and their corresponding tumor masks

We aimed to leverage these False Positive tumors' basic information as background knowledge, specifically using \method\ to generate similar tumors so that enhance model's generalization ability. Therefore, we need to construct the control conditions for tumor synthesis while descriptive text is required to contextualize each CT-tumor mask pair. A reference dataset with descriptive terminology is used as a foundational source. Through few-shot learning, GPT-4o is adapted to capture the language and visual features specific to tumor characterization.

Each False Positive tumor region is subsequently magnified and processed through GPT-4o, which generates descriptions aligned with the tumor's visual features based on a predefined set of high-frequency descriptive terms. This descriptive text is then paired with the zoomed-in mask and a healthy CT scan, forming a structured input for the diffusion model. This approach enables targeted augmentation, enriching the model's training data for improved tumor detection.

This approach focuses on augmenting the dataset with tumors that challenge current detection methods, aiming to improve model performance, specifically targeting an increase in the DSC by 4.7\% and sensitivity by 9.1\% for large kidney tumors as shown in \tableautorefname~\ref{sup_Ablation_Study1} and \tableautorefname~\ref{sup_Ablation_Study2}. By generating challenging yet realistic tumor instances, we seek to refine the model's ability to accurately detect diverse tumor presentations, addressing limitations in current diagnostic applications.

\clearpage
\section{Dataset and Implementation Details}
\subsection{Dataset}\label{sec:supp_data}
For the Diffusion Model training, the dataset comprises 173 CT scans, including 98 liver, 31 pancreas, and 78 kidney scans. Each scan contains uncertain or lesion regions, and the ground truth annotations are derived from detailed radiology reports. To ensure reliable data quality, true positive cases were selected from prior evaluations of DiffTumor~\cite{chen2024towards}. This subset consists of 66 liver, 15 pancreas, and 60 kidney scans, totaling 141 CT scans.

In training the Diffusion Model, paired inputs are crucial. Each input typically includes three elements: an unhealthy CT scan, the corresponding descriptive report, and the associated tumor mask. However, acquiring a dataset containing all three elements simultaneously is particularly challenging in real-world scenarios due to limited availability of such comprehensive datasets. This leads to two alternative options for training data:

\smallskip\noindent\textbf{CT-Mask Pair Data:}
This dataset includes CT scans paired with their corresponding tumor masks. While this option ensures the availability of spatial tumor annotations, it lacks descriptive textual reports. Generating accurate descriptive words for a given CT scan is currently constrained by the absence of robust tools. Moreover, obtaining reliable ground truth descriptions for such data pairs is difficult without radiologist involvement or advanced text generation methods.

\smallskip\noindent\textbf{CT-Report Pair Data:}
To address the issue of missing tumor masks, segmentation tools like DiffTumor~\cite{chen2024towards} can generate tumor masks directly from CT scans, filling dataset gaps. Descriptive reports can serve as ground truth to evaluate the accuracy of these generated masks. For example, \tableautorefname~\ref{tab:supp_groundtruth} demonstrates how reports are converted into binary ground truth labels (0 or 1) to validate tumor masks, ensuring consistency between textual observations and spatial annotations.

\subsection{Implementation Details}

\noindent\textbf{Diffusion Model:}
In this study, we train the corresponding Diffusion Model specifically for tumors of three different abdominal organs. The data preprocessing carried out during the training phase is identical to the approach used for training the Autoencoder Model. We utilize the Adam optimizer with hyperparameters $\beta_1=0.9$ and $\beta_2=0.999$, a learning rate of 0.0001, and a batch size of 10 per GPU. The training is conducted on 4 A6000 GPUs for a week, over a total of 60,000 epochs.

\smallskip\noindent\textbf{Segmentation Model:}
The code for the Segmentation Model is implemented in Python using MONAI\footnote{Cardoso~\etal~\cite{cardoso2022monai}: \href{https://monai.io/}{https://monai.io/}}. The orientation of CT scans is adjusted according to specific axcodes. Each scan is resampled to achieve isotropic spacing of $1.0 \times 1.0 \times 1.0~\text{mm}^3$. The intensity of each scan is truncated to the range [$-175, 250$] and then linearly normalized to [0, 1].

During training, we randomly crop fixed-sized $96 \times 96 \times 96$ regions centered on either a foreground or background voxel, following a predefined ratio. The input patch is randomly rotated by $90^\circ$ with a probability of 0.1, and its intensity is shifted by 0.1 with a probability of 0.2. To avoid confusion between organs on the right and left sides, mirroring augmentation is not employed.

All models trained on both synthetic and real tumors are trained for 2,000 epochs. The base learning rate is set to 0.0002, with a batch size of 4. We adopt a linear warmup strategy and use a cosine annealing learning rate schedule. The Segmentation Model is trained on eight A6000 GPUs for a total of 4 days.

For details on the tumor synthesis process during Segmentation Model training, please refer to the provided code. For inference, we use a sliding window strategy with an overlap ratio of 0.75. To exclude tumor predictions that do not belong to the respective organs, we post-process the predictions of the Segmentation Models using pseudo-labels of organs obtained from previous work\footnote{Liu~\etal~\cite{liu2023clip,liu2024universal}: \href{https://github.com/ljwztc/CLIP-Driven-Universal-Model}{https://github.com/ljwztc/CLIP-Driven-Universal-Model}}.

\begin{table*}[ht]
\centering
\scriptsize
    \begin{tabular}    {p{0.08\linewidth}p{0.03\linewidth}p{0.04\linewidth}p{0.04\linewidth}>{\raggedright\arraybackslash}p{0.7\linewidth}}
    \toprule
    \textbf{ID} & \textbf{Liver} & \textbf{Pancreas} & \textbf{Kidney} & \textbf{Report} \\ 
    \midrule
    g9wxm1kPLU & 1 & 0 & 0 & 
    Liver: Post right hepatectomy. Numerous hyperenhancing liver lesions, with index lesions as above. No definite new lesions. \newline 
    Pancreas: Post Whipple procedure. Stable mild dilatation of the main pancreatic duct, measuring up to 4 mm. \newline 
    Kidneys: Unremarkable \\ 
    \midrule
    1J07NmUKTS & 1 &  0 &  0 & 
    Liver: Calcifications within the right liver may represent granulomas. Unchanged mild left adrenal nodularity. \\
    \midrule
    aS1CGuWccw & 1 & 0 & 0 & 
    Liver: Multiple hepatic hypodensities, which are new or increased in size compared to prior, for example measuring 9 mm in hepatic segment 2 (6/36) and 4 mm in segment 5/6 (6/51). \newline 
    Pancreas: Unremarkable \newline 
    Kidneys: Unremarkable \\ 
    \midrule
    ZpA1AE7Laf & 1 & 0 & 0 & 
    Liver: Multiple hepatic hypodensities, which are new or increased in size compared to prior, for example measuring 9 mm in hepatic segment 2 (6/36) and 4 mm in segment 5/6 (6/51). \newline 
    Pancreas: Unremarkable\newline 
    Kidneys: Unremarkable \\ 
    \midrule
    fCineePb6z & 1 & 0 & 0 & 
    Liver: Known metastatic lesions are not significant change for prior. For example:  segment 6: 2.4 x 1.3 cm lesion ([DATE]) measured 2.6 x 1.4 cm previously segment 2: 0.6 cm lesion ([DATE]) measures 0.9 cm previously  segment 7: 0.9 cm lesion (4/31) measured 0.9 cm previously. Additional scattered subcentimeter foci, some of which are new from prior for example in segment [DATE] ([DATE] in segment [DATE] (4/48). \newline 
    Pancreas: Unremarkable \newline 
    Kidneys: Unremarkable \\ 
    \midrule
    Irx9vyFo8u & 1 & 0 & 0 & 
    Liver: INDEX LESIONS (Restaging): AI1: Segment 2 : 1.6 x 1 cm (Se/Im [DATE]), previously 1.6 x 1.1 cm AI2: Segment [DATE] : 1.6 x 1.3 cm (Se/Im 2/30), previously 1.6 x 1.3 cm. \newline 
    Pancreas: Unremarkable \newline 
    Kidneys: Unremarkable \\ 
    \midrule
    BT1EMp3cXm & 1 & 0 & 1 & 
    Liver: Index lesions as above. Decreased size of multiple hypoattenuating hepatic lesions. No new hepatic lesions. \newline
    Pancreas: Unremarkable \newline 
    Kidneys: Unchanged bilateral cysts and subcentimeter hypodensities that are too small to characterize. \\ 
    \midrule
    XokcrXmKyn & 1 & 0 & 0 & 
    Liver: Numerous metastases throughout the liver, multiple of which are increased in size compared to [DATE], Mass in segment 7 of the liver abutting the inferior cava measures 5.1 x 4.9 cm (303/43), previously 4.9 x 4.5 cm. There is similar associated slight mass effect on the inferior vena cava. Mass in segment 5 measures 5.3 x 4.8 cm (303/56), previously 4.3 x 4.2 cm. Mass in segment 2 measures 3.4 x 3.0 cm (303/41), previously 3.1 x 2.9 cm. \newline 
    Pancreas: Unremarkable\newline 
    Kidneys: Unremarkable \\ 
    \midrule
    8TfcZajFaf & 1 & 0 & 1 & 
    Liver: As indexed above. Interval decrease in size of multiple hypoattenuating hepatic lesions.\newline
    Pancreas: Unremarkable \newline 
    Kidneys: Bilateral subcentimeter hypodensities too small to further characterize. Similar bilateral pelviectasis. \\ 
    \midrule
    DGZfKbMJC4 & 1 & 0 & 0 & 
    Liver: Status post partial right hepatectomy. Decreased numerous hyperenhancing liver lesions, indexed above. Few hypoenhancing lesions are newly conspicuous from prior, indexed above, with direct comparison difficult due to difference in contrast timing technique (current exam portal venous with prior exam late hepatic arterial).\newline
    Pancreas: Status post Whipple with unremarkable residual pancreas.\newline
    Kidneys: Unremarkable \\ 
    \midrule
    hbC4w9qEGJ & 1 & 0 & 0 & 
    Liver: Numerous hepatic metastases, new and increased since [DATE]. Difficult to discretely measure given confluent lesions. For example, there is a confluence of metastases spanning 7.2 x 4.7 cm (303:30) in segment 7. Patent hepatic and portal veins.\newline
    Pancreas: Unremarkable \newline
    Kidneys: Unremarkable \\ 
    \midrule
    bYaN3j0vJd & 1 & 0 & 1 & 
    Liver: Numerous peripherally hyperenhancing lesions in both hepatic lobes are decreased in size since [DATE]. See reference lesions above.\newline 
    Pancreas: Unremarkable\newline
    Kidneys: Status post left nephrectomy. Multiple small peripherally hyperenhancing right renal lesions are stable compared to recent prior, but increased in size compared to more remote prior studies. No hydronephrosis. \\ 
    \midrule
    o2t9oeEWfC & 1 & 0 & 0 & 
    Liver: There are multiple conglomerate metastasis throughout the liver which demonstrate mixed interval response with some increased in size, some decreased in size and some stable compared to the prior study. A representative mass in segment 7 of the liver abutting the inferior vena cava measures 4.9 x 4.5 cm on image 34 series 304, previously 5.6 x 5.6 cm. There has been interval decrease in associated mass effect on the inferior vena cava. A mass in segment 5 measures 4.3 x 4.2 cm on image 49 series 304, unchanged A mass in segment 2 on image 30 series 304 measures 3.1 x 2.9 cm, previously 2.6 x 1.9\newline
    Pancreas: Unremarkable\newline 
    Kidneys: Unremarkable \\ 
    \midrule
    qBAQV8his3 & 1 & 0 & 1 & 
    Liver: Slightly enlarged ill-defined hypoenhancing liver lesions. Segment 2 lesion measures 1.6 x 1.1 cm, previously 1.4 x 1.0 cm. Segment 7 lesion measures 1.6 x 1.3 cm, previously 1.4 x 1.4 cm.\newline 
    Pancreas: Unremarkable\newline 
    Kidneys: Benign cysts \\ 
    \midrule
    \end{tabular}
\caption{\textbf{Presence of tumors in the liver, pancreas, and kidneys.} We extracted the descriptions related to the liver, pancreas, and kidneys from the original report and summarized the presence of tumors in these organs, where (1) indicates the presence of a tumor and (0) indicates the absence of a tumor. 'n/a' indicates that the radiological report is not available.}
\label{tab:supp_groundtruth}
\end{table*}

\clearpage
\section{Generalizable Across Different Patient Demographics}\label{sec:Generalizable}
\begin{figure}[ht]
  \centering
  \begin{subfigure}[t]{\linewidth}
    \centering
    \includegraphics[width=0.78\linewidth]{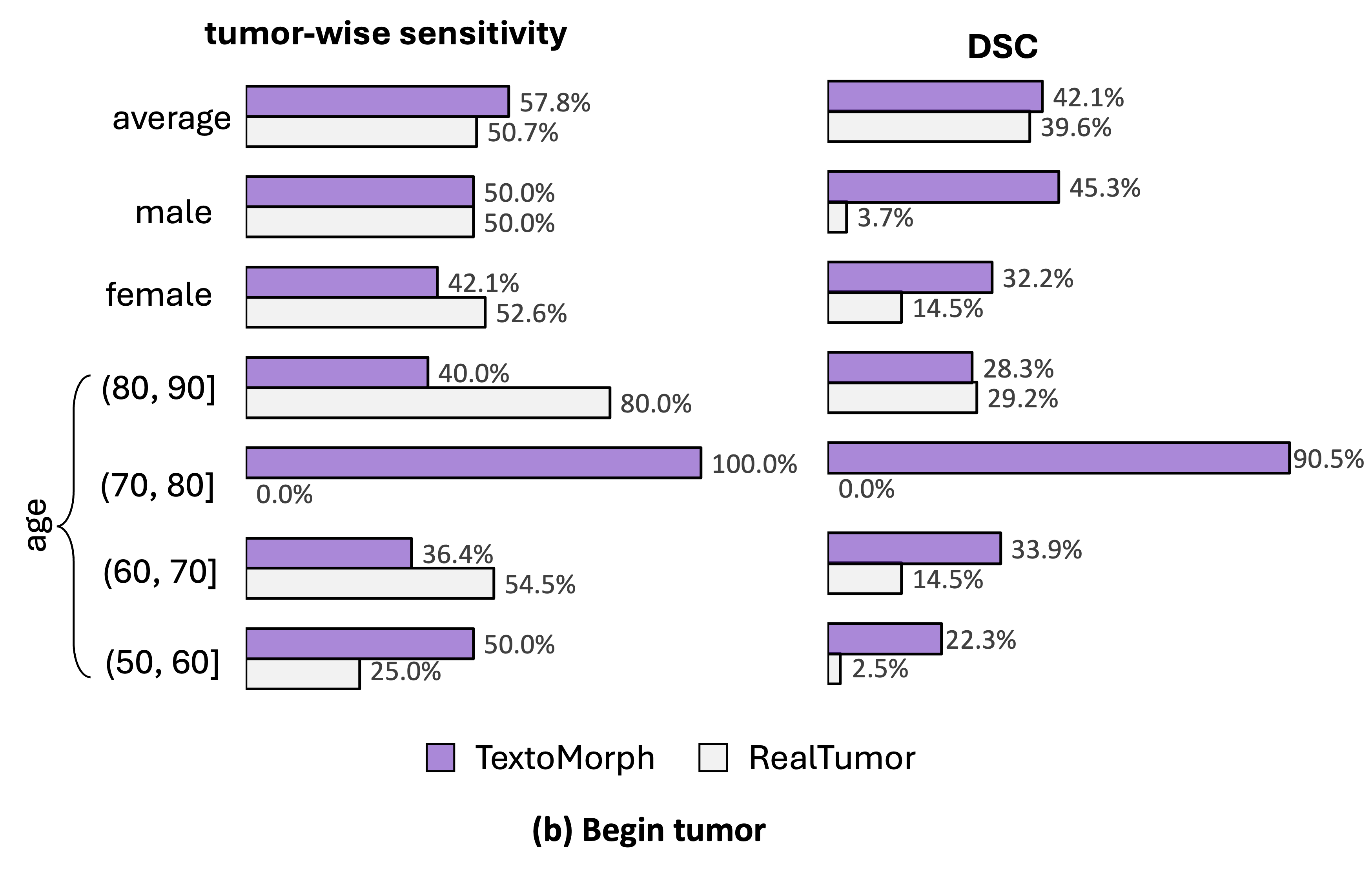}
    \label{fig:benign}
  \end{subfigure}
  
  \vspace{1em} 
  
  \begin{subfigure}[t]{\linewidth}
    \centering
    \includegraphics[width=0.78\linewidth]{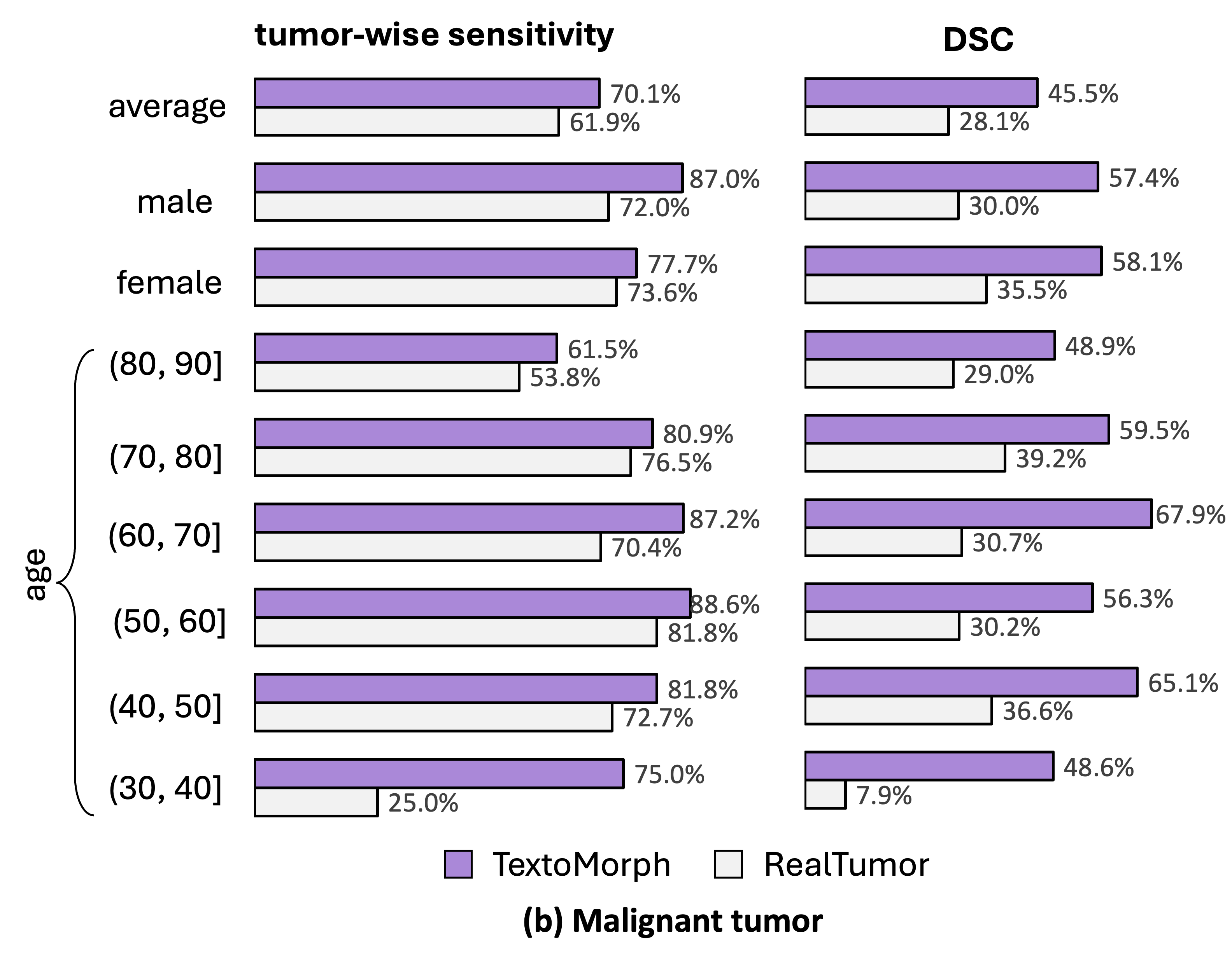}
    \label{fig:malignant}
  \end{subfigure}
  
  \caption{\textbf{Generalizable Across Different Patient Demographics.} A comparison of benign cysts and malignant tumors, illustrating the visual characteristics critical for tumor detection and segmentation. \method\ demonstrates consistent performance improvements in both tumor-wise Sensitivity (\%) and segmentation DSC (\%) across various patient groups.}
  \label{fig:comparison_vertical}
\end{figure}
\clearpage

\section{Descriptive Words and Descriptive Words Explanation}

\begin{table*}[ht]
\centering
\scriptsize
    \begin{tabular}{m{0.2\linewidth}|m{0.5\linewidth}|m{0.25\linewidth}}
    \hline
    \textbf{Descriptive Words} & \textbf{Explanation} & \textbf{Image} \\ \hline
    Hypoattenuating or Hypodense Lesions & These lesions appear as darker or lighter areas compared to the surrounding liver tissue, making them easily noticeable on the scan. They usually have smooth edges and uniform appearance. & \includegraphics[width=0.6\linewidth]{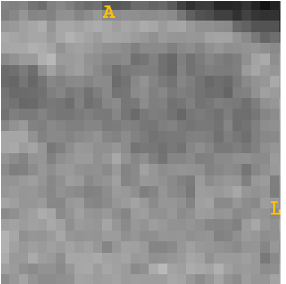} \\ \hline
    
    Enhancing and Washout & These lesions are bright and well-defined in the early phase of the scan, but their brightness fades over time, causing the edges to blur. This pattern makes them stand out in early and late scan phases.& \includegraphics[width=0.6\linewidth]{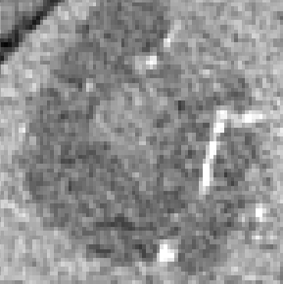} \\ \hline
    
    Cysts or Cystic Lesionså & These lesions are round or oval, with clear boundaries. They appear darker than the surrounding tissue, making them easy to identify as fluid-filled spaces. & \includegraphics[width=0.6\linewidth]{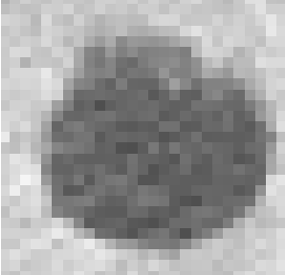} \\ \hline
    
    Heterogeneous or Mixed Enhancement & These lesions display areas with different levels of brightness or darkness within the same lesion. This uneven appearance shows complexity and can make the edges appear irregular. & \includegraphics[width=0.6\linewidth]{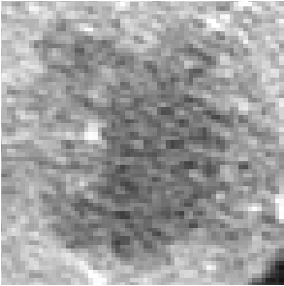} \\ \hline
    
    Fatty Infiltration or Steatosis & Fat deposits show up as large, lighter areas on the scan, either spread throughout the liver or concentrated in specific spots, giving the affected areas a more uniform, lighter tone. & \includegraphics[width=0.6\linewidth]{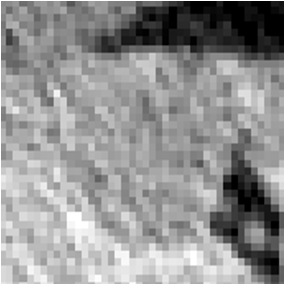} \\ \hline
    \end{tabular}
\caption{\textbf{Liver:} Descriptive words from liver radiology reports are paired with explanations and corresponding CT scan images to highlight their visual characteristics. Terms such as \textit{'Hypoattenuating Lesions'} and \textit{'Enhancing and Washout'} are explained in detail, focusing on features like texture, brightness, and shape.}
\end{table*}

\begin{table*}[ht]
\centering
\scriptsize
    \begin{tabular}{m{0.2\linewidth}|m{0.5\linewidth}|m{0.25\linewidth}}
    \hline
    \textbf{Descriptive Words} & \textbf{Explanation} & \textbf{Image} \\ \hline
    Hypoattenuating or Hypodense Lesions & These lesions appear as areas with lower density than the surrounding pancreas tissue, often depicted as darker regions on the scan. The lesions typically have smooth and distinct borders, creating a noticeable contrast with the adjacent tissues. & \includegraphics[width=0.6\linewidth]{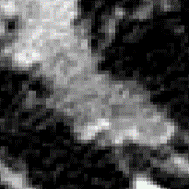} \\ \hline
    
    Atrophy and Calcifications & Atrophic areas are seen as regions with noticeable shrinkage of tissue, often accompanied by calcifications that appear as small, bright spots or patches. Calcifications are sharply defined, and their high contrast against the surrounding tissue makes them easily identifiable. & \includegraphics[width=0.6\linewidth]{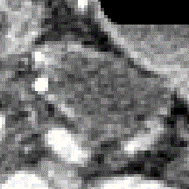} \\ \hline
    
    Ill-Defined or Poorly Defined Lesions & These lesions are irregular in shape, with fuzzy or blurry borders that blend into the surrounding tissue. The lack of clear boundaries makes them appear less distinct on the scan, often merging with the normal pancreas tissue in the image. & \includegraphics[width=0.6\linewidth]{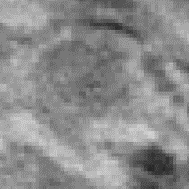} \\ \hline
    
    Necrotic Lesions & These lesions exhibit areas with varying brightness, indicating tissue death. The necrotic regions often have mixed density, with darker (dead tissue) and brighter (inflamed or surviving tissue) areas. The edges tend to be uneven and jagged, giving the lesion a complex and chaotic appearance. & \includegraphics[width=0.6\linewidth]{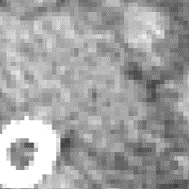} \\ \hline
    
    \end{tabular}
\caption{\textbf{Pancreas:} Key descriptive terms from pancreatic radiology reports are linked with detailed explanations and corresponding CT images to illustrate specific imaging characteristics. For instance, \textit{'Hypoattenuating Lesions'} appear as darker regions with distinct borders, while \textit{'Atrophy and Calcifications'} depict tissue shrinkage alongside bright, sharply defined calcifications. Additionally, \textit{'Ill-Defined Lesions'} feature irregular shapes with blurry edges blending into the surrounding tissue, and \textit{'Necrotic Lesions'} show mixed-density areas with uneven, jagged edges indicative of tissue death.}
\end{table*}

\begin{table*}[ht]
\centering
\label{tab:Descriptive_liver_tumor_vision}
\scriptsize
    \begin{tabular}{m{0.2\linewidth}|m{0.5\linewidth}|m{0.25\linewidth}}
    \hline
    \textbf{Descriptive Words} & \textbf{Explanation} & \textbf{Image} \\ \hline
    Hypoattenuating or Hypodense Lesions & These lesions appear as areas with reduced density compared to the surrounding kidney tissue, often appearing as darker regions on the scan. Their edges are typically smooth and well-defined, making them stand out against the normal tissue. & \includegraphics[width=0.6\linewidth]{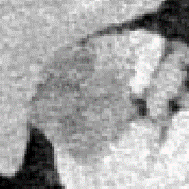} \\ \hline
    
    Enhancing or Heterogeneously Enhancing Masses & These masses exhibit a bright, high-contrast appearance during the early phases of the scan, with varied intensity across the lesion. Over time, the brightness may fade, leading to blurring of the edges. This dynamic contrast makes them visually distinct in both early and late phases. & \includegraphics[width=0.6\linewidth]{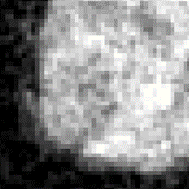} \\ \hline
    
    Renal Cysts & These cysts appear as smooth, round, or oval fluid-filled spaces with clear and sharp boundaries. They are typically darker than the surrounding tissue, making them easily distinguishable from solid kidney structures. & \includegraphics[width=0.6\linewidth]{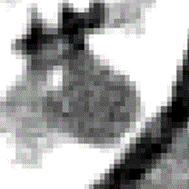} \\ \hline
    
    Renal Stones or Calculi & These appear as small, high-density spots on the scan due to their calcified nature. Their edges are sharp, and their brightness makes them stand out significantly against the lower-density surrounding tissue. & \includegraphics[width=0.6\linewidth]{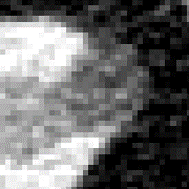} \\ \hline
    
    \end{tabular}
\caption{\textbf{Kidney:} Key descriptive terms from kidney radiology reports are paired with detailed explanations and representative CT images to highlight specific imaging characteristics. For example, \textit{'Hypoattenuating Lesions'} appear as darker regions with smooth and well-defined edges, while \textit{'Enhancing Masses'} show bright, high-contrast patterns with varying intensity across the lesion. \textit{'Renal Cysts'} are fluid-filled spaces with clear, sharp boundaries, and \textit{'Renal Stones'} are identified as small, high-density spots with sharp edges due to their calcified nature.}
\end{table*}

\begin{table*}[t]
\centering
\scriptsize
\begin{tabular}{l|p{0.9\linewidth}}
\hline
\textbf{Organ} & \textbf{Descriptive Words} \\ \hline
Liver & Hepatic cyst; Hypoattenuating hepatic lesion; Heterogeneous enhancement; Status post liver transplant; Cyst; Scattered hypodensities, likely cysts. \\
      & Cirrhosis; Focal fat along the falciform ligament; Hypodensities represent hemangiomas; Hypodensity, likely cysts; Focal fatty infiltration; Multiple cysts. \\
      & Ill-defined hypodensity; Hypoattenuating lesions; Cyst, mild hepatic steatosis; Lobulated hypodensity, possibly a cyst or hemangioma; Hepatic cysts, possible granulomas. \\
      & Arterially hyperenhancing lesions with washout; Isodense lesion with washout; Focus of arterial enhancement, indeterminate; Capsular retraction in metastases. \\
      & Scattered hypoattenuating lesions; Metastatic lesions; Hypodense lesions, consistent with hepatic metastases; Hyperenhancing and hypoenhancing liver lesions. \\ \hline
Pancreas & Hypoenhancing mass; Hypodense lesion, likely neoplasm; Atrophy, calcifications; Decreased enhancement; Hypoattenuating, infiltrative. \\
         & Necrotic, involving arteries; Ductal dilation; Ill-defined, atrophy; Focal mass, atrophy; Hypoattenuating, ductal dilation. \\
         & Ill-defined, duct dilation; Fat stranding; Ill-defined, hypointense. \\ \hline
Kidney & Symmetric renal cortical enhancement; No hydronephrosis; Renal cysts, scattered hypodensities; Atrophic kidneys, likely cysts; Nonobstructing stone. \\
       & Hypodensity, likely benign; Angiomyolipoma; Scattered renal cysts; Renal cysts, hypoattenuating foci; Nonobstructive renal calculi. \\
       & Atrophic kidney, stent in place; Renal cyst, hypodensities; Renal cortical atrophy, renal cysts; Low-density lesions, likely cysts. \\
       & Enhancing renal mass; Hypoenhancing renal mass, cystic lesion; Exophytic mass, tumor thrombus; Heterogeneously enhancing exophytic mass. \\
       & Partially calcified mass; Benign cysts; Renal mass. \\ \hline
\end{tabular}
\caption{Consolidated Radiology Descriptions Across Liver, Pancreas, and Kidney}
\end{table*}

\end{document}